\documentclass[pra,showpacs,twocolumn,aps,superscriptaddress,a4paper,longbibliography]{revtex4-1}
\usepackage{dcolumn,amssymb,amsmath,amsfonts,graphicx,latexsym,color}
\usepackage[utf8]{inputenc}
\usepackage[T1]{fontenc}
\usepackage{epstopdf}

\usepackage{ulem}
\usepackage{epstopdf}
\usepackage{subfigure}
\usepackage{float}
\usepackage{mathrsfs}
\usepackage{bbm}
\usepackage{psfrag}

\usepackage{verbatim}
\usepackage{multirow}
\usepackage{diagbox}
\usepackage[colorlinks,citecolor=blue]{hyperref}

\usepackage{CJK}
\begin{document}


\title{Light-induced half-quantized Hall effect and axion insulator}

\author{Fang Qin}
\email{qinfang@nus.edu.sg}
\affiliation{Department of Physics, National University of Singapore, Singapore 117551, Singapore}

\author{Ching Hua Lee}
\email{phylch@nus.edu.sg}
\affiliation{Department of Physics, National University of Singapore, Singapore 117551, Singapore}

\author{Rui Chen}
\email{chenr@hubu.edu.cn}
\affiliation{Department of Physics, Hubei University, Wuhan 430062, China}


\begin{abstract}
Motivated by the recent experimental realization of the half-quantized Hall effect phase in a three-dimensional (3D) semi-magnetic topological insulator [\href{https://www.nature.com/articles/s41567-021-01490-y}{M. Mogi et al., Nature Physics \textbf{18}, 390 (2022)}], we propose a scheme for realizing the half-quantized Hall effect and axion insulator in experimentally mature 3D topological insulator heterostructures. Our approach involves optically pumping and/or magnetically doping the topological insulator surface, such as to break time reversal and gap out the Dirac cones. By toggling between left and right circularly polarized optical pumping, the sign of the half-integer Hall conductance from each of the surface Dirac cones can be controlled, such as to yield half-quantized ($0+1/2$), axion ($-1/2+1/2=0$), and Chern ($1/2+1/2=1$) insulator phases. We substantiate our results based on detailed band structure and Berry curvature numerics on the Floquet Hamiltonian in the high-frequency limit. Our paper showcases how topological phases can be obtained through mature experimental approaches such as magnetic layer doping and circularly polarized laser pumping and opens up potential device applications such as a polarization chirality-controlled topological transistor.
\end{abstract}

\maketitle

\section{Introduction}\label{1}
The Hall conductance is a paradigmatic example of a directly measurable topological invariant~\cite{haldane1988model}. Taking only integer values in purely (two-dimensional) 2D band insulators, it can interestingly assume half-integer values when the 2D system is a surface of a 3D topological insulator, such as (Bi,Sb)$_{2}$Te$_{3}$~\cite{zhang2009topological,chang2013experimental,mogi2022experimental} and Bi$_2$Se$_3$~\cite{zhang2009topological}. Lately, this has attracted a lot of attention in the context of magnetically doped~\cite{nenno2020Axion,mogi2017tailoring,mogi2017magnetic,xiao2018realization,chang2013experimental,yoshimi2015quantum,okugawa2022correlated} and intrinsic antiferromagnetic topological insulators~\cite{liu2020robust,qiu2023Axion,deng2020quantum,li2021nonlocal}, where local time-reversal symmetry breaking~\cite{qi2008topological,nomura2011surface,qi2011topological,fu2007topological,gu2016holographic,lee2014lattice,haldane1988model} opens up a gap in the surface Dirac cone and gives rise to a half-quantized surface Hall conductance~\cite{mogi2022experimental,varnava2018surfaces,lee2018electromagnetic,liu2020anisotropic,fijalkowski2021any,li2021quantized,xu2019higher,chen2021using,chen2023side,chen2023halfquantized,zou2022half,zou2023half,zhou2022transport,ning2023robustness,gu2021spectral,nenno2020Axion,mogi2017tailoring,mogi2017magnetic,xiao2018realization,chu2011surface,konig2014half}.

In a key recent experiment, half-quantized Hall conductance has been observed in a 3D semi-magnetic topological insulator~\cite{mogi2022experimental}, where one surface state is gapped by magnetic doping and the opposite surface is non-magnetic and gapless. By physically gapping the Dirac cone in a 3D topological insulator, this mechanism for the half-quantized Hall effect is not just a vivid manifestation of the parity anomaly~\cite{mogi2022experimental,zou2022half,zhou2022transport,zou2023half,chen2023halfquantized,ning2023robustness}, but also provides a route towards other coveted and closely related topological phases, such as the axion insulator and the Chern insulator. 
The axion insulator phase, characterized by a zero Hall plateau~\cite{mogi2017tailoring,mogi2017magnetic,xiao2018realization} accompanied by a quantized topological magnetoelectric effect~\cite{wang2015quantized,morimoto2015topological,qi2009inducing,maciejko2010topological,tse2010giant,yu2019magnetic,sekine2021axion}, has been realized in a 3D sandwich heterostructure involving magnetic topological insulator layers. The Chern insulator, also known as the quantum anomalous Hall insulator~\cite{haldane1988model,yu2010quantized,lu2010massive,shan2010effective,lu2013quantum,chang2013experimental,jotzu2014experimental}, has also been realized in an intrinsic magnetic topological insulator~\cite{chang2013experimental} in the absence of an external magnetic field.

Transitions between these topological phases are crucially controlled by the gap of the surface Dirac cones of the 3D topological insulators. One extremely versatile avenue for selectively inducing gaps in surface Dirac cones is via Floquet driving with circularly polarized light, an established approach already known for driving a variety of topological transitions~\cite{bukov2015universal,eckardt2015high,chen2018floquet1,chen2018floquet2,du2022weyl,wang2022floquet,qin2022light,qin2022phase,dabiri2021light,dabiri2021engineering,dabiri2022floquet,askarpour2022light,pervishko2018impact,zhu2023floquet,nag2021anomalous,ghosh2022systematic}. Indeed, it has been experimentally demonstrated that circularly polarized light can gap out the helical Dirac cones of 3D topological insulators~\cite{wang2013observation,mahmood2016selective} by breaking time-reversal symmetry. This light-induced anomalous Hall effect has also been experimentally observed in graphene~\cite{mciver2020light}. So far, the existing literature has only focused on how circularly polarized light can induce a Chern insulator phase in a 3D topological insulator -- whether it can also induce the half-quantized Hall effect phase and the axion insulator remains an open question.

\begin{figure}[htpb]
\centering
\includegraphics[width=0.48\textwidth]{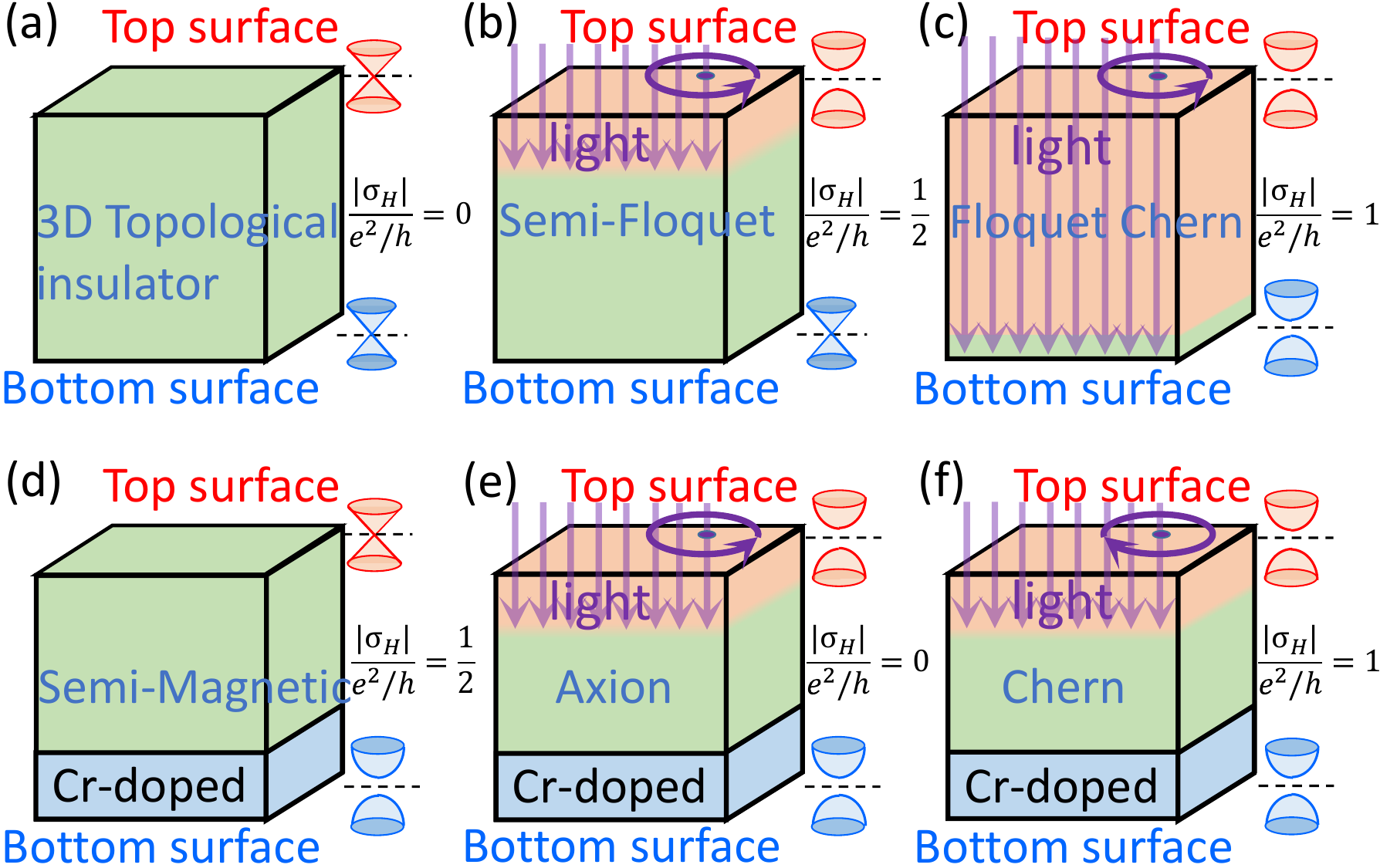}
\caption{(Color online) Schematic of various topological phases that result when a 3D topological insulator (Bi,Sb)$_{2}$Te$_{3}$ or Bi$_2$Se$_3$ is optically driven from above, and/or doped with magnetic Cr in its bottom layers.
(a) Ordinary 3D $\mathbb{Z}_2$ topological insulator without Cr doping and optical pumping, with two opposite gapless surface Dirac cones. (b) When irradiated with light that penetrates only the top layer (orange), the top surface Dirac cone becomes gapped due to time-reversal breaking, resulting in a semi-Floquet topological insulator (Floquet-induced half-quantized Hall effect). (c) Floquet-induced Chern phase, where the radiation penetrates both layers (orange) and gaps out both Dirac cones. (d) When optical driving is absent, but Cr doping (blue) breaks time reversal in the bottom layers, gapping out its Dirac cone, a semi-magnetic topological insulator (magnetic doping-induced half-quantized Hall effect) results. [(e),(f)] When there is optical driving upon one surface and Cr-doping in the other, both surfaces have Dirac cones gapped out, but their chiralities can be controlled. Depending on whether left-handed (e) or right-handed (f) circularly polarized light (purple circular arrow) is used, we obtain either (e) an axion insulator with zero net Chern number or (f) an effective Chern phase with nonzero Hall conductivity.}  \label{Fig:Schematic}
\end{figure}

Below, with close reference to Fig.~\ref{Fig:Schematic}, we provide a pedagogical summary of these closely-related topological phases, such that the precise roles of optical driving and magnetic doping in this paper are made clear. Note that the total Hall conductivity is directly related to the number of gapped surface Dirac cones: Every Dirac cone contributes a Hall conductivity of $\pm\frac{e^2}{4\pi\hbar}$, depending on the chirality of the gap~\cite{mogi2022experimental,qi2008topological,nomura2011surface,qi2011topological,fu2007topological,gu2016holographic,morimoto2015topological,li2019intrinsic}. 

\begin{itemize}
\item Figure~\ref{Fig:Schematic}(a): 3D $\mathbb{Z}_2$ topological insulator phase, i.e., when the 3D topological insulator (Bi,Sb)$_{2}$Te$_{3}$ or Bi$_2$Se$_3$ is without Cr doping and light pumping, the top and bottom surfaces possess opposite Dirac cones of opposite chirality due to time-reversal symmetry, and they together contribute zero Hall conductance. 

\item Figure~\ref{Fig:Schematic}(b): Semi-Floquet topological insulator phase (Floquet-induced half-quantized Hall effect), i.e., when the Dirac cone gap is opened by optical driving at the top but not the bottom surface, resulting in a half-quantized Hall conductance. This can happen when the topological insulator is subjected to optical pumping at a sufficiently weak intensity such that only the top surface is irradiated (orange), while the bottom surface is shielded by the skin effect~\cite{humlivcek2014raman,hada2016bandgap}. 

\item Figure~\ref{Fig:Schematic}(c): Floquet-induced Chern topological insulator phase, i.e., when the Dirac cones at both top and bottom surfaces are gapped by the time-reversal breaking from circularly polarized light (purple circular arrow), resulting in an integer-quantized total Hall conductance, such that the two surfaces together constitute a Chern insulator. This occurs when the radiation is sufficiently strong such that it passes through the sample (orange), reaching the bottom surface.

\item Figure~\ref{Fig:Schematic}(d): Semi-magnetic topological insulator phase~\cite{mogi2022experimental} (magnetic doping induced half-quantized Hall effect phase), i.e., when (magnetic) Cr doping (blue) is selectively introduced only in the bottom layers of the sample, such that the Dirac cone becomes gapped only in the bottom surface due to broken time-reversal symmetry. The unpaired, gapped Dirac cone contributes a half-quantized Hall conductance. 

\item Figure~\ref{Fig:Schematic}(e): Axion insulator phase~\cite{mogi2017tailoring,mogi2017magnetic,xiao2018realization,zhu2022axionic} (semi-magnetic Floquet axion insulator phase), examined in detail in this paper. The topological insulator is both radiated with left-handed circularly polarized light that only penetrates the top layers (orange), and is Cr-doped in the bottom layers (blue). This can gap the Dirac cones in both the top and bottom surfaces, albeit with opposite chiralities, resulting in a zero total quantum Hall conductance.

\item Figure~\ref{Fig:Schematic}(f): We dub this the semi-magnetic Floquet Chern topological insulator phase, which is examined in detail in this paper. Like in Fig.~\ref{Fig:Schematic}(e), we subject the topological insulator to optical pumping in the top layer (orange) and Cr doping (blue) in the bottom layer, but with the optical polarization being right-handed instead of left handed (clockwise purple circular arrow). This produces Dirac cones with the same chirality in both top and bottom layers and gaps them out to result in half-quantized Hall conductance contributions of the same sign. The top and bottom surfaces thus combine to form a Chern insulator.
\end{itemize}

In this paper, we provide a quantitative study of how the combination of optical driving and Cr magnetic doping can induce all of the above-mentioned phases in realistic thin samples of (Bi,Sb)$_{2}$Te$_{3}$ and Bi$_2$Se$_3$, particularly the semi-Floquet topological insulator and the axion insulator. The main results are as follows: First, by adjusting the intensity of the circularly polarized pumped light, the penetration depth can be adjusted such that either one or both surfaces of the topological insulator are irradiated, an approach not studied in previous works on Floquet driving with topological insulators~\cite{bukov2015universal,eckardt2015high,chen2018floquet1,chen2018floquet2,du2022weyl,wang2022floquet,qin2022light,qin2022phase,dabiri2021light,dabiri2021engineering,dabiri2022floquet,askarpour2022light,pervishko2018impact,zhu2023floquet,wang2013observation,mahmood2016selective}.  
Second, our approach lends a new way to easily switch between the axion (zero Hall plateau) and the Chern insulator (quantized Hall plateau) phases by reversing the optical polarization, thus complementing existing experimental efforts in realizing the semi-magnetic, axion~\cite{mogi2017tailoring,mogi2017magnetic,xiao2018realization}, and Chern~\cite{yu2010quantized,chang2013experimental,jotzu2014experimental,mciver2020light} topological insulator phases as well as suggesting potential technological applications such as chirality-selective topological transistors~\cite{sun2023magnetic}.

This paper is organized as follows: In Section~\ref{2}, we introduce the lattice Hamiltonian for our TI heterostructure, as well as that for the magnetic doping. In Section~\ref{3}, we derive the corresponding Floquet effective Hamiltonian for the optical driving in the high-frequency limit. In Section~\ref{4}, we present the energy dispersions and Hall conductance with fixed laser penetration depth by numerically diagonalizing the corresponding Floquet tight-binding Hamiltonian.
In Section~\ref{5}, we substantiate our Hall conductance results by presenting the actual spatial distributions of the surface Dirac bands as well as their Berry curvature profiles. 
In Section~\ref{6}, we further discuss related alternative routes for realizing our Floquet axion and Chern phases without the use of magnetic doping.

\section{Model}\label{2}

We begin by writing down the tight-binding model Hamiltonian for 3D topological insulators (Bi,Sb)$_{2}$Te$_{3}$ and Bi$_2$Se$_3$, which is given by~\cite{zou2023half,chu2011surface,zhang2009topological,liu2010model,ding2020hinged} 
\begin{align}\label{eq:H0}
H^{(0)}({\bf k}) 
\!=\!&\sum_{j_{z}}\left[C^{\dagger}_{{\bf k},j_{z}}h({\bf k})C_{{\bf k},j_{z}}+ C^{\dagger}_{{\bf k},j_{z}}T_{z}C_{{\bf k},j_{z}+1} \right.\nonumber\\
&\left. + C^{\dagger}_{{\bf k},j_{z}+1}T_{z}^{\dagger}C_{{\bf k},j_{z}} \right],
\end{align} where 
\begin{align} \label{eq:H0kk}
h({\bf k})&\!=\!\left[m_{0}-2t_{z}-4t_{||}\left(\sin^{2}\frac{k_{x}a}{2}+\sin^{2}\frac{k_{y}a}{2}\right) \right]\sigma_{0}\otimes\tau_{z} \nonumber\\
&~~\!+\! \lambda_{||}\sin(k_{x}a)\sigma_{x}\otimes\tau_{x} \!+\! \lambda_{||}\sin(k_{y}a)\sigma_{y}\otimes\tau_{x},\\
T_{z}&\!=\!t_{z}\sigma_{0}\otimes\tau_{z}\!-\! i\frac{\lambda_{z}}{2}\sigma_{z}\otimes\tau_{x},
\end{align} 
where $a$ is the lattice constant along $x$ or $y$ direction, $C^{\dagger}_{{\bf k},j_{z}}$ and $C_{{\bf k},j_{z}}$ are the four-component creation and annihilation operators at position $j_{z}$ along $z$ direction with wave vector ${\bf k}=(k_{x},k_{y})$, $\sigma_{x,y,z}$ and $\tau_{x,y,z}$ are the Pauli matrices for the spin and orbital degrees of freedom, respectively, $\sigma_{0}$ ($\tau_{0}$) is a $2\times2$ unit matrix, $a$ is the lattice constant, $m_0$, $t_{z}$, $t_{||}$, $\lambda_{z}$, and $\lambda_{||}$ are model parameters. The basis $(|1p_{z}^{+},\uparrow\rangle , |2p_{z}^{-},\uparrow\rangle , |1p_{z}^{+},\downarrow\rangle , |2p_{z}^{-},\downarrow\rangle )$ are the hybridized states of Te or Se $p_{z}$ orbital (1) and Sb or Bi $p_{z}$ orbital (2), with even ($+$) and odd ($-$) parities, up ($\uparrow$) and down ($\downarrow$) spins~\cite{yu2010quantized,lu2010massive,shan2010effective}. The material parameters for both (Bi,Sb)$_{2}$Te$_{3}$ and Bi$_2$Se$_3$ are similar~\cite{zhang2009topological,chang2013experimental,mogi2022experimental,zou2023half}: $t_{z}=0.40$ eV, $t_{||}=0.566$ eV, $\lambda_{z}=0.44$ eV, $\lambda_{||}=0.41$ eV, and $m_{0}=0.28$ eV. The lattice constant is $a=1$ nm in the $x$-$y$ directions and $a_z=0.5$ nm in the $z$ direction; we have kept the latter in real-space form so as to implement the top and bottom surfaces. The detailed derivations for the momentum-space and real-space tight-binding models  Eqs.~\eqref{eq:H0kk} and \eqref{eq:H0} can be found in Appendix \ref{Appendix_A} and \ref{Appendix_B} respectively.

If we consider Cr doping such as to introduce additional time-reversal breaking at the bottom layer, the additional exchange field Hamiltonian from the magnetic dopants reads~\cite{yu2010quantized,lu2013quantum,chen2019effects,dabiri2021light,dabiri2021engineering,qin2022phase}
\begin{align}\label{eq:Hd}
\Delta H_{d} = \sum_{j_{z}}C^{\dagger}_{{\bf k},j_{z}}\left[V_{z}(j_{z})\sigma_{z}\otimes\tau_{0}\right]C_{{\bf k},j_{z}},
\end{align}
where $V_{z}(j_{z})$ is the magnitude of the bulk magnetic moment~\cite{chang2013experimental}, which is only nonzero on the lattice sites of the bottom layers $j_{z}=n_{z}-1,n_{z}$ where $n_{z}$ is the total number of layers, i.e., we have $V_{z}(j_{z})=V_{z}=0.1$ eV~\cite{chang2013experimental,mogi2022experimental,zou2023half} at $j_{z}=n_{z}-1,n_{z}$ and $V_{z}(j_{z})=0$ elsewhere. The corresponding matrix form of the real-space tight-binding model with Cr doping can be found in Appendix \ref{Appendix_C}.

Explicitly, we see that the magnetic doping Hamiltonian \eqref{eq:Hd} breaks time-reversal symmetry from ${\cal T}[H^{(0)}({\bf k})+\Delta H_{d}]{\cal T}^{-1}\!\neq\![H^{(0)}(-{\bf k})+\Delta H_{d}]$, with the time-reversal operator being ${\cal T}\!=\!\sigma_{y}\otimes\tau_{0}{\cal K} $~\cite{schindler2018higher}, ${\cal K}$ the complex conjugation operator. The corresponding detailed derivations can be found in Appendix \ref{Appendix_D}.

\section{Floquet Hamiltonian}\label{3}

We next describe the optical driving field and how it leads to the effective Floquet Hamiltonian, starting from our model above. The optical driving field propagating along the $z$ direction in our topological insulator (Bi,Sb)$_{2}$Te$_{3}$ or Bi$_2$Se$_3$ can be expressed as ${\bf E}(z,t) = \partial{\bf A}(z,t)/\partial t = E(z)( \cos(\omega t),  \cos(\omega t + \varphi), 0 )$, where $E(z)=E_{0}e^{-z/\delta}$ is the amplitude of the optical field, $z$ is the position along the $z$ direction, and $\delta$ is the skin penetration depth due to optical absorption, as given by $\delta=\sqrt{2\rho/(\omega\mu)}\sqrt{\sqrt{1+(\rho\omega\varepsilon)^{2}}+\rho\omega\varepsilon}$~\cite{vander2006rf,jordan1968electromagnetic} where $\rho$ is the resistivity of the bulk material, $\omega$ is the angular frequency of the applied light beam, $\mu=\mu_{r}\mu_{0}$ is the permeability of the bulk material with the relative magnetic permeability $\mu_{r}$ and the vacuum permeability $\mu_{0}$, $\varepsilon=\varepsilon_{r}\varepsilon_{0}$ is the permittivity of the bulk material, with the relative permittivity $\varepsilon_{r}$ and the vacuum permittivity $\varepsilon_{0}$. For Bi$_2$Se$_3$, the conductivity is $\sigma\sim8.8\times10^4$ S/m~\cite{bauer2021surface,brom2012structural,yin2017plasmonics}, i.e., $\rho=1/\sigma\sim1.136\times10^{-5}$ m/S. For Bi$_2$Te$_3$, the conductivity is $\sigma\sim1.22\times10^5$ S/m~\cite{park2016thermal}, i.e., $\rho=1/\sigma\sim8.197\times10^{-6}$ m/S. 
From the above, we have ${\bf A}(z,t)={\bf A}(z,t+T)=\omega^{-1}E(z)( \sin(\omega t),  \sin(\omega t + \varphi), 0 )$ which is of period $T=2\pi/\omega$, $\omega$ being the optical frequency. The phase delay $\varphi=\mp\pi/2$ introduces left- or right-handed circular polarization.
Since we are interested in the off-resonant regime in which the central Floquet band is far away from other replicas, such that the high-frequency expansion is applicable, we set the driving frequency in this paper as $\hbar\omega=3.82$ eV ($\omega\sim5.80\times10^{3}$ THz that is in the deep ultraviolet), which is much larger than the bandwidth~\cite{qin2022light,qin2022phase,dabiri2021light,dabiri2021engineering,dabiri2022floquet,askarpour2022light,pervishko2018impact}. 
For Bi$_2$Se$_3$, the skin penetration depth is set as $\delta\sim16.3$ nm at $\hbar\omega=3.82$ eV based on experimental data~\cite{humlivcek2014raman}; for Bi$_2$Te$_3$, it is $24.6$ nm~\cite{humlivcek2014raman,hada2016bandgap} at the same frequency.

Under optical driving, the motion of lattice electrons is governed by minimal substitution of the lattice momentum with the electromagnetic gauge field ${\bf A}(z,t)$, i.e., the Peierls substitution $t_{||}\to t_{||}\exp\left[i\frac{e}{\hbar}\int_{{\bf r}_{j}}^{{\bf r}_{j'}}{\bf A}(z,t)\cdot d{\bf r}\right]$ and $\lambda_{||}\to \lambda_{||}\exp\left[i\frac{e}{\hbar}\int_{{\bf r}_{j}}^{{\bf r}_{j'}}{\bf A}(z,t)\cdot d{\bf r}\right]$, where ${\bf r}_{j}$ is the coordinate of the lattice site $j$, $j\rq{}=j\pm1$, $-e$ is the electron charge, and $\hbar$ is the reduced Planck's constant.
Hence, upon irradiation with light, the photon-dressed effective Hamiltonian is given by
\begin{align}\label{eq:Ht}
H({\bf k},t)= H^{(0)}\left({\bf k} - \frac{e}{\hbar}{\bf A}(z,t)\right)+\Delta H_{d}.
\end{align} 

We next derive the effective static Floquet Hamiltonian~\cite{oka2009photovoltaic,calvo2015floquet,seshadri2022engineering,seshadri2019generating,seshadri2022floquet,lee2018floquet,lee2021quenched,zheng2014floquet} $H^{(F)}({\bf k})=\frac{i}{T}\ln\left[\mathcal{T}e^{-i\int^T_0 H({\bf k},t)dt}\right]$ with the periodic driving ``averaged'' over through $\mathcal{T}$, the time-ordering operator. In the high-frequency regime, a closed-form solution exists via the Magnus expansion~\cite{magnus1954exponential,blanes2009magnus,lee2018floquet,zheng2014floquet,bukov2015universal,eckardt2015high,chen2018floquet1,chen2018floquet2,du2022weyl,wang2022floquet,qin2022light,qin2022phase} \begin{align}\label{eq:HF0} 
H^{(F)}({\bf k}) = H_{0} + \sum_{n=1}^{\infty}\frac{[H_{-n}, H_{n}]}{n\hbar\omega}+{\cal O}(\omega^{-2}),
\end{align} where $H_{m-m'} = \frac{1}{T} \int_{0}^{T}H^{(0)}({\bf k},t) e^{i(m-m')\omega t}dt$ with $m$ and $m'$ as integers. The concrete analytical expressions for $H_{0}$, $H_{-n}$, and $H_{n}$ can be found in Appendix \ref{Appendix_E}.
From Eq.~\eqref{eq:HF0}, the Floquet Hamiltonian can be evaluated as
\begin{widetext}\begin{align}
H^{(F)}({\bf k}) 
&\!=\!\sum_{j_z}\!\left\{\!m_{0}-2t_{z}-4t_{||}+2{\cal J}_{0}(A(z)a)t_{||}\left[\cos(k_{x}a)+\cos(k_{y}a)\right]\!\right\}\!C_{{\bf k},j_z}^{\dagger}\gamma_{0}C_{{\bf k},j_z} \nonumber\\
&\!+\!t_{z}\sum_{j_z}\!\left(\!C_{{\bf k},j_z}^{\dagger}\gamma_{0}C_{{\bf k},j_z+1} + C_{{\bf k},j_z+1}^{\dagger}\gamma_{0}C_{{\bf k},j_z} \!\right)\! 
\!-\! i\frac{\lambda_{z}}{2}\sum_{j_z}\!\left(\!C_{{\bf k},j_z}^{\dagger}\gamma_{z}C_{{\bf k},j_z+1}\!-\! C_{{\bf k},j_z+1}^{\dagger}\gamma_{z}C_{{\bf k},j_z}\!\right)\! \nonumber\\
& \!+\! {\cal J}_{0}(A(z)a)\lambda_{||}\sin(k_{x}a)\sum_{j_z}C_{{\bf k},j_z}^{\dagger}\gamma_{x}C_{{\bf k},j_z} \!+\! {\cal J}_{0}(A(z)a)\lambda_{||}\sin(k_{y}a)\sum_{j_z}C_{{\bf k},j_z}^{\dagger}\gamma_{y}C_{{\bf k},j_z} \nonumber\\
&\!+\!\sum_{j_z}C_{{\bf k},j_z}^{\dagger}\sum_{n\in{\rm odd},n>0}\frac{2i\lambda_{||}{\cal J}_{n}^{2}(A(z)a)}{n\hbar\omega}\sin\left(n\varphi\right)
\left\{ 2t_{||}\cos(k_{x}a)\sin(k_{y}a)[\gamma_{x}, \gamma_{0}] 
+2t_{||}\sin(k_{x}a)\cos(k_{y}a)[\gamma_{0}, \gamma_{y}] \right.\nonumber\\
&\left. - \lambda_{||}\cos(k_{x}a)\cos(k_{y}a)[\gamma_{x}, \gamma_{y}] \right\}C_{{\bf k},j_z},
\label{eq:H_F}
\end{align}\end{widetext} where ${\cal J}_{n}(A(z)a)$ is the $n$th Bessel function of the first kind~\cite{temme1996special}, $\gamma_{0}=\sigma_{0}\otimes\tau_{z}$, $\gamma_{j=x,y,z}=\sigma_{j=x,y,z}\otimes\tau_{x}$, and we used $\varphi=\pi/2$.
The detailed matrix form of the tight-binding Floquet Hamiltonian can be found in Appendix \ref{Appendix_F}. 
Here $A(z)=A_{0}e^{-z/\delta}$ with $A_{0}=eE_{0}/(\hbar\omega)$. Additionally, the validity of the high-frequency expansion can be found in Appendix \ref{Appendix_G}.

Note that it should not be taken for granted that optical driving will simply induce the Chern insulator phase: when $\varphi=0$, we find that the Floquet Hamiltonian ${\cal T}H^{(F)}$ (\ref{eq:H_F}) still satisfies time-reversal symmetry, i.e., ${\cal T}H^{(F)}({\bf k}){\cal T}^{-1}\!=\!H^{(F)}(-{\bf k})$ with the time-reversal symmetry operator being ${\cal T}\!=\!\sigma_{y}\otimes\tau_{0}{\cal K} $~\cite{schindler2018higher}, where ${\cal K}$ is the complex conjugation operator. 
However, when $\varphi=\frac{\pi}{2}$, the terms containing $\varphi$ in $H^{(F)}({\bf k})$ [Eq.~\eqref{eq:H_F}] lead to ${\cal T}H^{(F)}({\bf k}){\cal T}^{-1}\!\neq\!H^{(F)}(-{\bf k})$, which breaks time-reversal symmetry. The corresponding derivation details can be found in Appendix \ref{Appendix_H}.

\section{Energy spectra and Hall conductances}\label{4}

\begin{figure}
\centering
\includegraphics[width=0.43\textwidth]{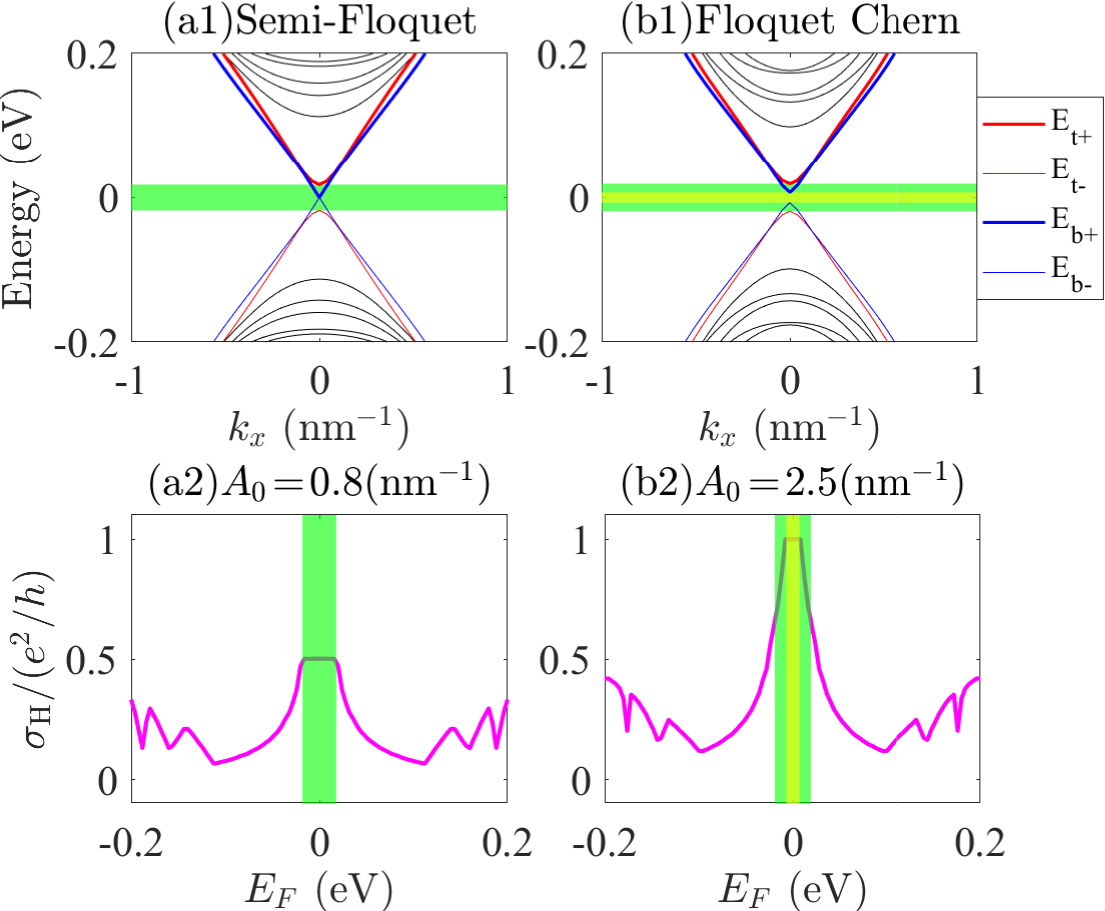}
\caption{(Color online)
Floquet band structures and their corresponding Hall conductivity for the Floquet optical-driven Hamiltonian~\eqref{eq:H_F} of Bi$_2$Se$_3$ without Cr doping. [(a1),(b1)] Floquet band structures under open boundary conditions along the $z$ direction and periodic boundary conditions along the $x$ and $y$ directions.
In (a1), $A_{0}=0.8$ nm$^{-1}$ and the driving radiation only gaps out the top surface Dirac cone (red). 
In (b1), $A_{0}=2.5$ nm$^{-1}$ and the driving radiation penetrates both surfaces, gapping both the top and bottom surface Dirac cones (red and blue).
Here, $E_{t\pm}$ and $E_{b\pm}$ are the energies of the top and bottom surface bands; the subscripts ``$\pm$'' respectively denote the lowest conduction band or highest valence band. 
The green and yellow shaded intervals indicate the band widths of the top and bottom surfaces, respectively. 
[(a2),(b2)] Hall conductance as a function of the Fermi energy $E_{F}$, corresponding to the light intensities in (a1) $A_{0}=0.8$ nm$^{-1}$ and (b1) $A_{0}=2.5$ nm$^{-1}$. They exhibit half-integer (semi-Floquet) and integer-quantized (Chern-Floquet) Hall conductivity in the gap, respectively.
The other parameters are $k_y=0$, sample thickness $L_z=30$ nm, $a_{z}=0.5$ nm, $a=1$ nm, $t_{z}=0.40$ eV, $t_{||}=0.566$ eV, $\lambda_{z}=0.44$ eV, $\lambda_{||}=0.41$ eV, $m_{0}=0.28$ eV, $\delta=16.3$ nm, $\hbar\omega=3.82$ eV, and $\varphi=-\pi/2$.
}
\label{Fig:E_C_Vz0_A_Lz30_delta163_together}
\end{figure}

\begin{figure}[htpb]
\centering
\includegraphics[width=.5\textwidth]{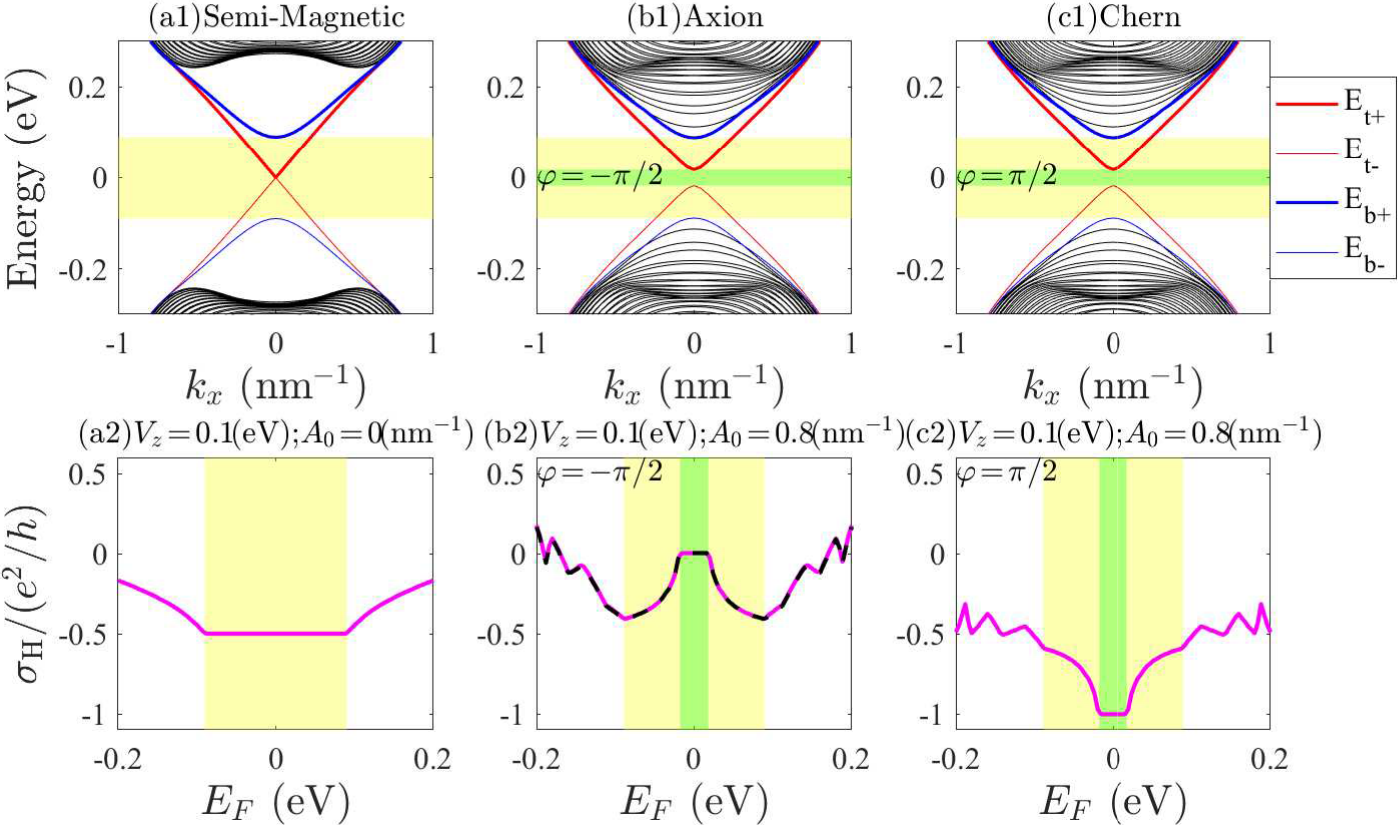}
\caption{(Color online) Band and Floquet band structures and their corresponding Hall conductivity for the Hamiltonian of Bi$_2$Se$_3$ with Cr doping. [(a1)-(c1)] Energy bands for the tight-binding Hamiltonian of the Bi$_2$Se$_3$ under open boundary conditions along the $z$ direction and periodic boundary conditions along the $x$ and $y$ directions.
In (a1), $V_z=0.1$ eV, $A_{0}=0$ nm$^{-1}$, and the magnetic doping gaps out the bottom surface Dirac cone (blue). 
In (b1), $V_z=0.1$ eV, $A_{0}=0.8$ nm$^{-1}$, $\varphi=-\pi/2$, and the driving radiation penetrates the top surface, gapping the top surface Dirac cone (red). The magnetic doping gaps out the bottom surface Dirac cone (blue).
In (c1), $V_z=0.1$ eV, $A_{0}=0.8$ nm$^{-1}$, $\varphi=\pi/2$, and the driving radiation penetrates the top surface, gapping the top surface Dirac cone (red). The magnetic doping gaps out the bottom surface Dirac cone (blue).
Here, $E_{t\pm}$ and $E_{b\pm}$ are the energies of the top and bottom surface bands, subscripts ``$\pm$'' respectively denote the lowest conduction band or highest valence band. 
The green and yellow shaded intervals indicate the band widths of the top and bottom surfaces, respectively. 
[(a2)-(c2)] Hall conductance as a function of the Fermi energy $E_{F}$, corresponding to the parameters in (a1) $V_z=0.1$ eV, $A_{0}=0$ nm$^{-1}$; (b1) $V_z=0.1$ eV, $A_{0}=0.8$ nm$^{-1}$, $\varphi=-\pi/2$; and (c1) $V_z=0.1$ eV, $A_{0}=0.8$ nm$^{-1}$, $\varphi=\pi/2$. They exhibit half-integer (semi-magnetic), zero (axion), and integer-quantized (Chern) Hall conductivity in the gap, respectively.  
The thickness of the Cr-doped layer is $d=1$ nm from the bottom surface, and the laser penetration depth is $\delta=16.3$ nm under $\hbar\omega=3.82$ eV~\cite{humlivcek2014raman} accompanying the light incoming from the top surface. 
The black dashed line in (b2) is the combination of the Hall conductances of Fig.~\ref{Fig:E_C_Vz0_A_Lz30_delta163_together}(a2) and Fig.~\ref{Fig:E_C_Lz30_delta163_inverse_together}(a2).
The other parameters are the same as those in Fig.~\ref{Fig:E_C_Vz0_A_Lz30_delta163_together}.
} \label{Fig:E_C_Lz30_delta163_inverse_together}
\end{figure}

To calculate the Hall conductance of the system, we first define the Hall conductance as~\cite{sun2020analytical,otrokov2019unique}
\begin{align}
\sigma_{H}=\frac{e^2}{h}\frac{1}{2\pi}\sum_{j}\int f(E_{j}-E_{F})\Omega_{j,z}(k_x,k_y)dk_{x}dk_{y},
\end{align} where $f(E_{j}-E_{F})=\Theta(E_{F}-E_{j})$ is the Fermi distribution function in the zero-temperature limit, $\Theta(x)$ the Heaviside function~\cite{qin2015three,qin2018high,qin2019polaron,qin2018universal,qin2017width}, $E_{F}$ is the Fermi energy, and $\Omega_{j,z}(k_x,k_y)$ is the Berry curvature for the energy band $j$, which reads~\cite{shen2017topological,thouless1982quantized,fukui2005chern,sticlet2013distant,chen2023halfquantized,chen2018floquet1}
\begin{align}
\Omega_{j,z}(k_x,k_y)
\!=\!-2{\rm Im}\sum_{i\neq j}\frac{\langle j|(\partial{\cal H}/\partial k_x)|i\rangle\langle i|(\partial{\cal H}/\partial k_y)|j\rangle}{(E_{j} - E_{i})^{2}}.\label{eq:Berry}
\end{align} 
Here $\cal{H}$ and its eigenvectors $|i\rangle$, $|j\rangle$ are taken to be those of the Floquet Hamiltonian for cases with optical pumping.
The detailed derivation for the Berry curvature \eqref{eq:Berry} can be found in Appendix \ref{Appendix_I}.

We next present the energy spectra and how the presence of each gapped surface Dirac cone leads to a half-quantized Hall conductivity. First, we consider cases with only optical driving and no Cr doping. As shown in Fig.~\ref{Fig:E_C_Vz0_A_Lz30_delta163_together}(a1) for the semi-Floquet case from Fig.~\ref{Fig:Schematic}(b), Bi$_2$Se$_3$ is under optical pumping from the top surface with a weak light intensity of $A_{0}=0.8$ nm$^{-1}$, such that only the top surface (red curve) opens up a gap and the bottom surface (blue curve) is gapless. Only the gapped top surface contributes a half-quantized Hall conductance within its gap (green), as shown in Fig.~\ref{Fig:E_C_Vz0_A_Lz30_delta163_together}(a2). 
In Fig.~\ref{Fig:E_C_Vz0_A_Lz30_delta163_together}(b1) for the Floquet Chern case from Fig.~\ref{Fig:Schematic}(c), Bi$_2$Se$_3$ is under optical pumping from the top surface with a strong light intensity of $A_{0}=2.5$ nm$^{-1}$, such that it penetrates both the top and bottom surfaces and gaps out their Dirac cones, which together contribute a quantized Hall conductance, as shown in the gapped region of both surfaces (yellow) in Fig.~\ref{Fig:E_C_Vz0_A_Lz30_delta163_together}(b2).

We next discuss cases where magnetic Cr doping $\Delta H_d$ is added to the bottom layers of the topological insulator, as presented in Fig.~\ref{Fig:E_C_Lz30_delta163_inverse_together}.  
Shown in Fig.~\ref{Fig:E_C_Lz30_delta163_inverse_together}(a1) is the band structure without any Floquet driving. Since Cr doping is only present at the bottom, the bottom surface Dirac cone (blue) gaps out, resulting in a negative half-quantized Hall conductance within the gap (yellow) [Fig.~\ref{Fig:E_C_Lz30_delta163_inverse_together}(a2)]. This is the semi-magnetic topological insulator phase from Fig.~\ref{Fig:Schematic}(d). 
Shown in Figs.~\ref{Fig:E_C_Lz30_delta163_inverse_together}(b) and \ref{Fig:E_C_Lz30_delta163_inverse_together}(c) are cases where the top surfaces are additionally Floquet-irradiated by left-handed and right-handed circularly polarized light, respectively. Both the top (red) and bottom (blue) surface Dirac cones are gapped -- the top by optical driving and the bottom by magnetic doping. While each contributes a half-quantized Hall conductance, in (b) with left-handed polarization, the Dirac cones are of opposite chirality, resulting in opposite Hall conductances that cancel [Fig.~\ref{Fig:E_C_Lz30_delta163_inverse_together}(b2)]. In Fig.~\ref{Fig:E_C_Lz30_delta163_inverse_together}(c) with right-handed polarization, the top surface Dirac cone's chirality is flipped, giving rise to an integer quantized Hall conductance within the gap (green) [Fig.~\ref{Fig:E_C_Lz30_delta163_inverse_together}(c2)]. These are respectively the cases (e) and (f) from Fig.~\ref{Fig:Schematic}, namely the semi-magnetic Floquet axion and Chern insulators.

\section{State density profile and Berry curvature distribution}\label{5}

In order to more precisely trace the origin of the Hall conductivity from the surface band structures, we present the detailed distribution of the surface bands in real space as well as their Berry curvatures in momentum space.

\begin{figure}[htpb]
\centering
\includegraphics[width=0.48\textwidth]{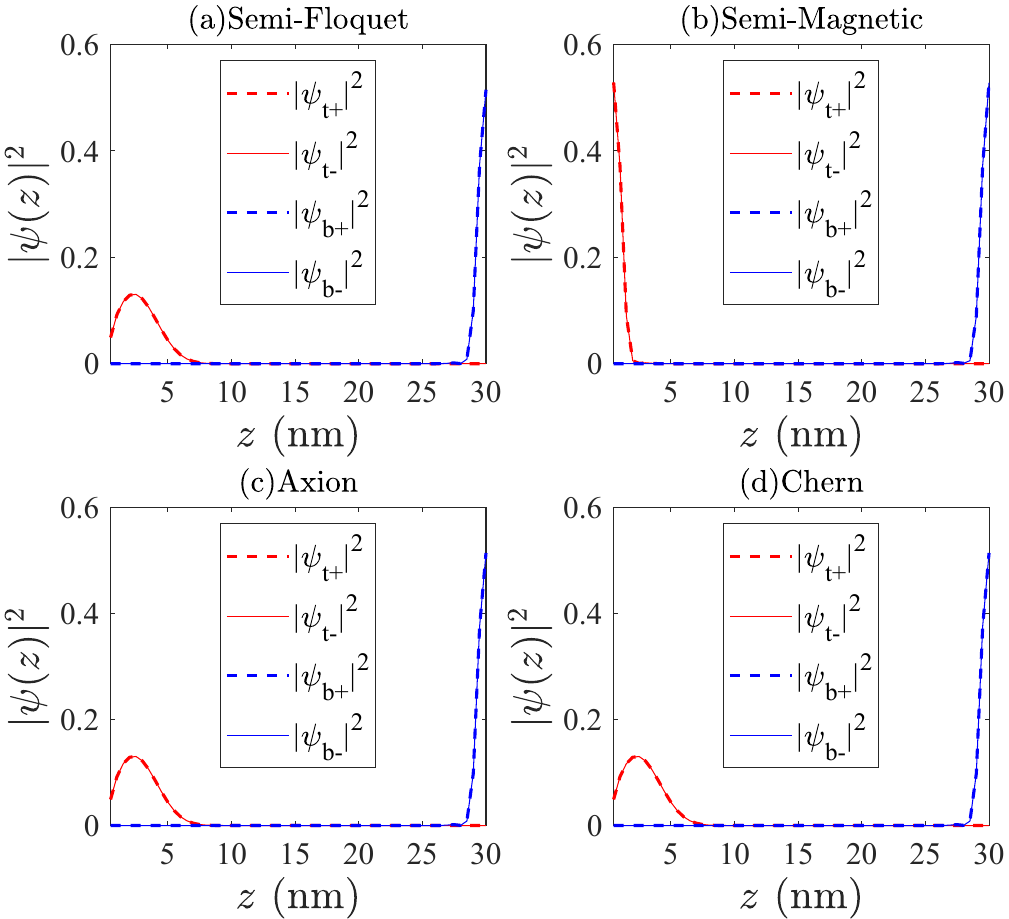}
\caption{(Color online) Spatial state distribution $|\psi(z)|^{2}$ along the vertical direction $z$ for the lowest conduction (subscript ``+\rq\rq) and highest valence (subscript ``-\rq\rq) bands of the top surface state ($|\psi_{t\pm}|^{2}$ -- red lines) and bottom surface state ($|\psi_{b\pm}|^{2}$ -- blue lines) at $k_{x}=k_{y}=0$. Indeed, the supposed top and bottom surface states are concentrated near the top ($z=0.5$ nm) and bottom $z=30$ nm boundaries. 
(a) Semi-Floquet topological insulator with $V_z=0$ eV, $A_{0}=0.8$ nm$^{-1}$, and $\varphi=-\pi/2$. (b) Semi-magnetic topological insulator with $V_z=0.1$ eV and $A_{0}=0$ nm$^{-1}$. (c) Axion insulator with $V_z=0.1$ eV, $A_{0}=0.8$ nm$^{-1}$, and $\varphi=-\pi/2$. (d) Chern topological insulator with $V_z=0.1$ eV, $A_{0}=0.8$ nm$^{-1}$, and $\varphi=\pi/2$. 
The other parameters are the same as those in Fig.~\ref{Fig:E_C_Lz30_delta163_inverse_together}.}  \label{Fig:P_Lz30_delta163_together}
\end{figure}

\begin{figure}[htpb]
\centering
\includegraphics[width=0.5\textwidth]{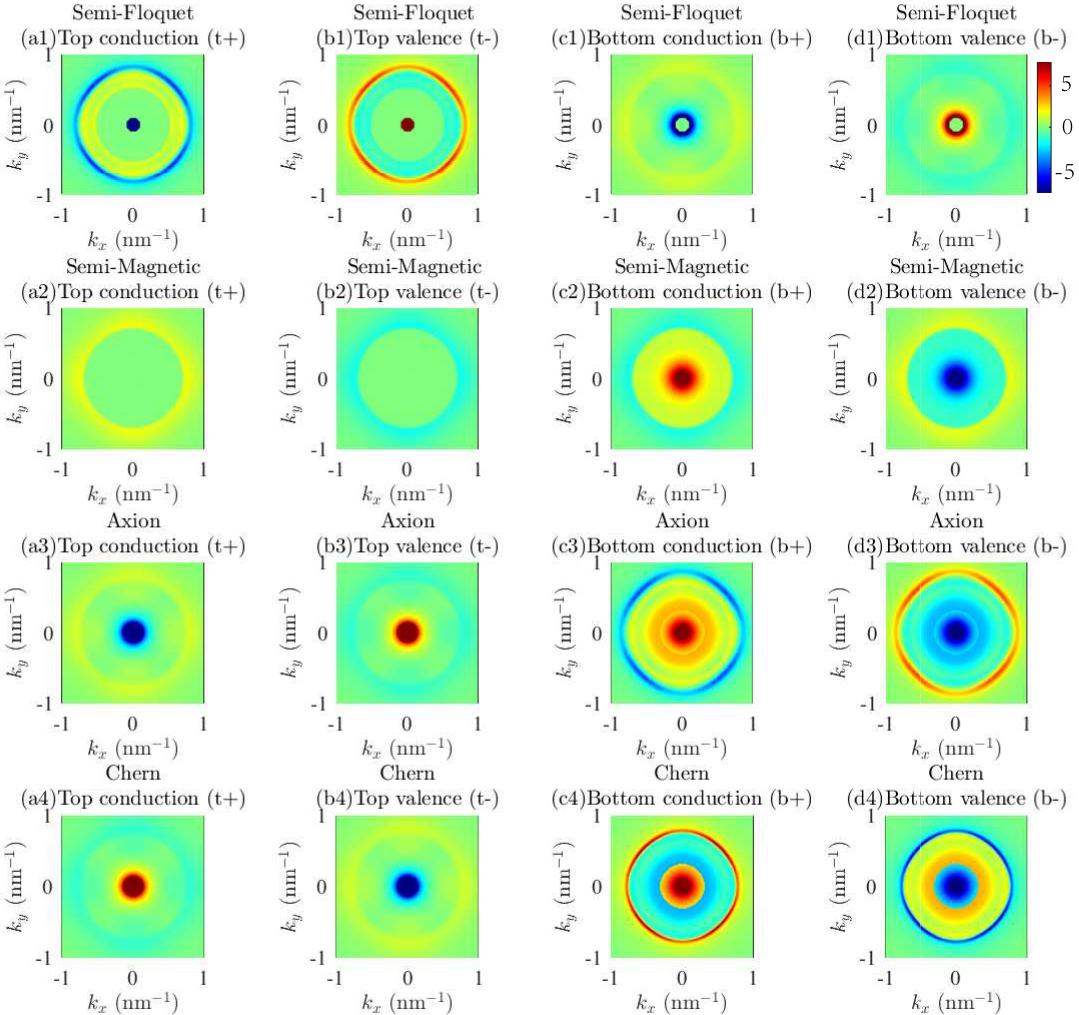}
\caption{(Color online) Berry curvature distributions for the lowest conduction (``+\rq\rq) and highest valence bands (``-\rq\rq) of the top (t) and bottom (b) surfaces, for the four scenarios from Fig.~\ref{Fig:P_Lz30_delta163_together}. [(a1)-(d1)] Semi-Floquet topological insulator with $V_z=0$ eV, $A_{0}=0.8$ nm$^{-1}$, and $\varphi=-\pi/2$. [(a2)-(d2)] Semi-magnetic topological insulator with $V_z=0.1$ eV and $A_{0}=0$ nm$^{-1}$. [(a3)-(d3)] Axion insulator with $V_z=0.1$ eV, $A_{0}=0.8$ nm$^{-1}$, and $\varphi=-\pi/2$. [(a4)-(d4)] Chern topological insulator with $V_z=0.1$ eV, $A_{0}=0.8$ nm$^{-1}$, and $\varphi=\pi/2$, which is equal but opposite to that of (c) due to the reversed polarization of the optical driving. We see Berry curvature peaks around $k_x=k_y=0$ when gapped Dirac cones exist, and another ring of peaks at larger $k_x,k_y$ when the bands merge into the bulk.
The other parameters are the same as those in Fig.~\ref{Fig:E_C_Lz30_delta163_inverse_together}.
}  \label{Fig:B_Lz30_together}
\end{figure}

Figure~\ref{Fig:P_Lz30_delta163_together} shows the spatial state distributions $|\psi(z)|^{2}$ of the four most relevant surface bands ($|\psi_{t\pm}|^{2}$ -- red lines and $|\psi_{b\pm}|^{2}$ -- blue lines) in the cases shown in Figs.~\ref{Fig:Schematic}(b), \ref{Fig:Schematic}(d), \ref{Fig:Schematic}(e), and \ref{Fig:Schematic}(f). As evident in Figs.~2 and 3, these bands are the ones closest to the gap, as labeled by subscripts ``$t+$'' (top surface, conduction) and ``$b+$'' (bottom surface, conduction), ``$t-$'' (top surface, valence), ``$b-$'' (bottom surface, valence). These states are plotted for $k_{x}=k_{y}=0$ where the Dirac cones, if any, reside.

From Fig.~\ref{Fig:P_Lz30_delta163_together}, it is evident that in all cases, the putative top (t) and bottom (b) states are indeed localized near the top surface (small $z$) and bottom surface (large $z$), respectively, with identical distributions for corresponding conduction and valence bands. The non-Floquet case in Fig.~\ref{Fig:P_Lz30_delta163_together}(b) has the most localized Dirac cone states, but the top surface state (red) in the other three Floquet cases [Figs.~\ref{Fig:P_Lz30_delta163_together}(a), \ref{Fig:P_Lz30_delta163_together}(c), and \ref{Fig:P_Lz30_delta163_together}(d)] are still sufficiently localized in the top 10 layers ($z=5$ nm), such that they would be greatly affected by circularly polarized pumping light incident on the top surface.

In Fig.~\ref{Fig:B_Lz30_together}, we present the Berry curvature distributions for the lowest conduction (subscript ``+\rq\rq) and highest valence (subscript ``-\rq\rq) bands of both the top (t) and bottom (b) surfaces, for each of the four cases shown in Figs.~\ref{Fig:Schematic}(d), \ref{Fig:Schematic}(b), \ref{Fig:Schematic}(e), and \ref{Fig:Schematic}(f). Due to the approximate rotational symmetry of Dirac cones near the gapless point in the Brillouin zone, the Berry curvature distributions in Fig.~\ref{Fig:B_Lz30_together} also look approximately rotation-invariant. In most of these plots, there is a ring of peak Berry curvature at around 0.8 nm$^{-1}$ when the surface bands merge into the bulk. But more importantly, at small $k_x,k_y$, we see even more intense Berry curvature contributions whenever there are gapped Dirac cones; when the Dirac cone is gapless, the Berry curvature disappears.

As shown in Figs.~\ref{Fig:B_Lz30_together}(a1)--\ref{Fig:B_Lz30_together}(d1), i.e., the semi-Floquet topological insulator phase, due to the light propagating vertically only from the top surface, the Berry curvature is mainly distributed in the center of the Brillouin zone for the gapped top surface state. But there is almost no distribution in the center of the Brillouin zone for the bottom gapless surface state without magnetic doping.
As shown in Figs.~\ref{Fig:B_Lz30_together}(a2)--\ref{Fig:B_Lz30_together}(d2), i.e., the semi-magnetic topological insulator phase, due to the magnetic doping only in the last two layers at the bottom, the Berry curvature is mainly distributed in the center of the Brillouin zone for the bottom surface Dirac cone. But there is almost no distribution in the center of the Brillouin zone $k_x=k_y=0$ for the top surface state, which is gapless. 
In Figs.~\ref{Fig:B_Lz30_together}(a3)--\ref{Fig:B_Lz30_together}(d3) and Figs.~\ref{Fig:B_Lz30_together}(a4)--\ref{Fig:B_Lz30_together}(d4), the Berry curvature is always concentrated at the center of the Brillouin zone because both surfaces are gapped -- from the irradiation on the top surface and the magnetic doping in the last two layers at the bottom. Comparing Fig.~\ref{Fig:B_Lz30_together}(a3) with Fig.~\ref{Fig:B_Lz30_together}(a4) for the lowest conduction band of the top surface state, one can find that the sign of the Berry curvature is opposite due to the opposite chirality of the light polarization. The same conclusion can be found by comparing Fig.~\ref{Fig:B_Lz30_together}(b3) with Fig.~\ref{Fig:B_Lz30_together}(b4) for the highest valence band of the top surface state.

\section{Alternative route towards Floquet Axion and Chern insulators}\label{6}
 
We briefly discuss an alternative route to achieve the Floquet axion and Chern insulators without the use of magnetic doping. The idea is to irradiate both the top and bottom surfaces of the topological insulator sample simultaneously, so as to break time reversal on both surfaces purely through Floquet driving. By using two different lasers (instead of one laser beam that must pass through the sample), both surfaces can be independently driven, and there is also no need for a beam that is sufficiently strong for full penetration.

\begin{figure}[htpb]
\centering
\includegraphics[width=0.4\textwidth]{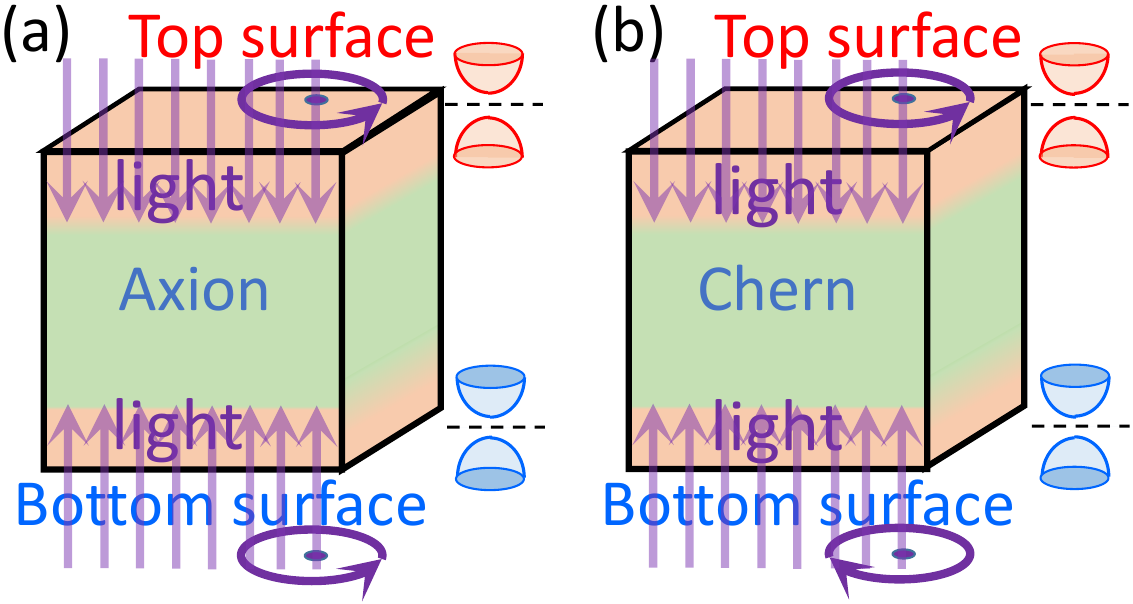}
\caption{(Color online) A 3D topological insulator sample of arbitrary thickness and without magnetic doping can be induced into either the Floquet axion or Chern phases by means of two lasers, one upon each surface. (a) Floquet axion insulator under a left-handed circularly polarized laser incoming from the top surface and a right-handed circularly polarized light incoming from the bottom surface. (b) Chern insulator phase under two left-handed circularly polarized lights incoming from the top and bottom surfaces, respectively. }  \label{Fig:Schematic_light}
\end{figure}

As described in Figs.~\ref{Fig:Schematic_light}(a) and \ref{Fig:Schematic_light}(b), when we shine on both surfaces simultaneously with two laser beams propagating in opposite directions (without any magnetic doping), there are two possible topological phases. One is (a) the Floquet axion insulator phase (the two different circularly polarized lights have opposite directions of polarization; for example, one is a left-handed circularly polarized light and the other is a right-handed circularly polarized light). The other one is (b) the Floquet Chern insulator phase (the two different circularly polarized lights have the same direction of polarization; for example, they are both left-handed circularly polarized lights).

The relative independence of tuning the top and bottom surfaces' lasers can potentially be useful for a chirality-controlled topological transistor, reminiscent of~\cite{sun2023magnetic}. The \lq\lq{}on\rq\rq{} state [quantized conductance -- Fig.~\ref{Fig:Schematic_light}(b)] and the \lq\lq{}off\rq\rq{} state [zero conductance -- Fig.~\ref{Fig:Schematic_light}(a)] can be easily toggled by changing the chirality of either circularly polarized laser. If ultrafast switching between the polarization directions can be performed, our setup may even serve as a platform for a Floquet quench involving Chern and axion phases, whose interplay can be explored in future work.

\subsection{Axion-Chern quench}\label{}

\begin{figure}[htpb]
\centering
\includegraphics[width=.5\textwidth]{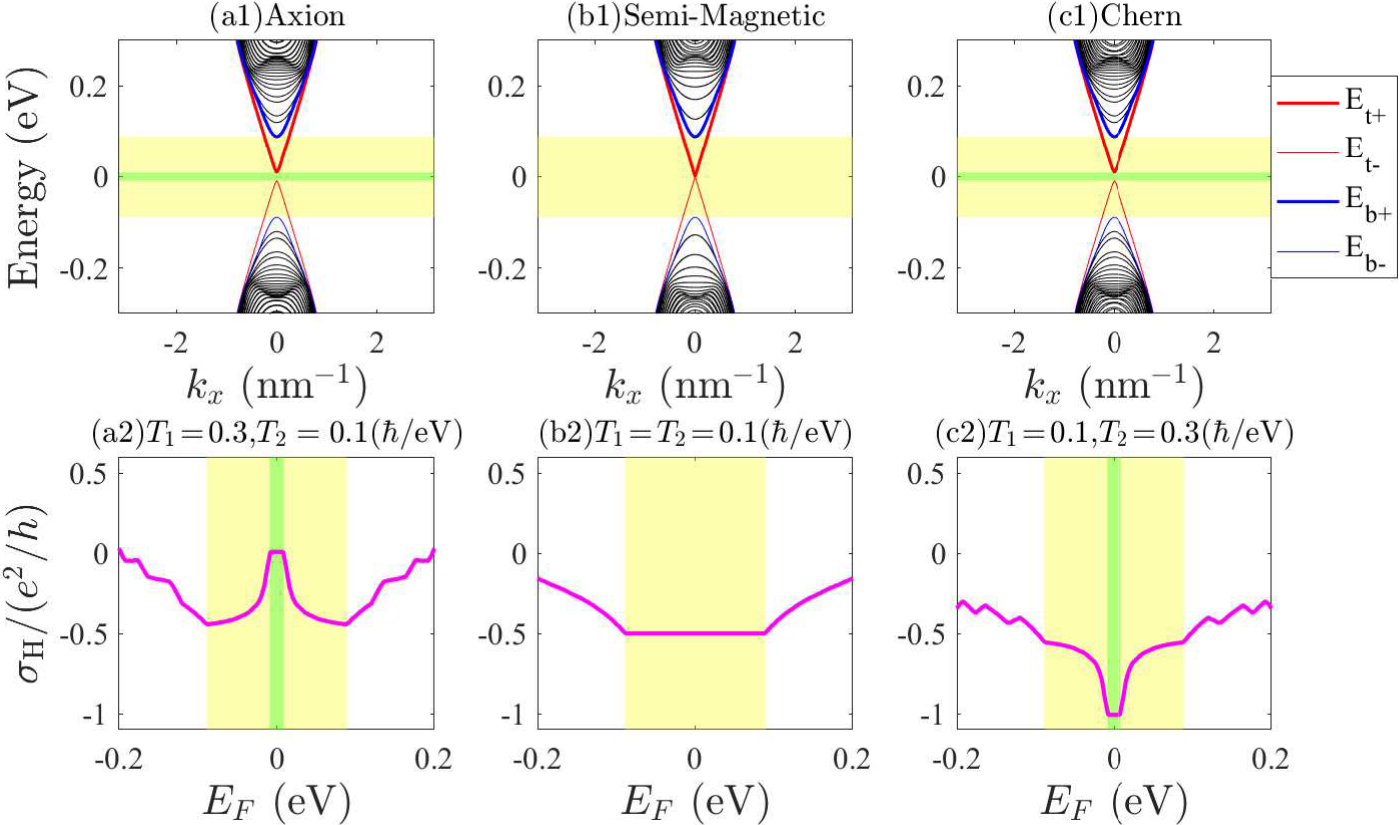}
\caption{(Color online) Floquet band structures and their corresponding Hall conductivity for the effective Hamiltonian \eqref{eq:Heff} under quench dynamics. [(a1)-(c1)] Floquet energy bands.
In (a1) $T_{1}=0.3$ $\hbar/$eV, $T_{2}=0.1$ $\hbar/$eV, and the magnetic doping gaps out the bottom surface Dirac cone (blue). The ${\cal H}_{\rm L}({\bf k})$ is dominant, gapping the top surface Dirac cone (red). 
In (b1), $T_{1}=T_{2}=0.1$ $\hbar/$eV, and the magnetic doping gaps out the bottom surface Dirac cone (blue). The contributions from ${\cal H}_{\rm L}({\bf k})$ and ${\cal H}_{\rm R}({\bf k})$ cancel each other out, so that the top surface is gapless (red).
In (c1), $T_{1}=0.1$ $\hbar/$eV, $T_{2}=0.3$ $\hbar/$eV, and the magnetic doping gaps out the bottom surface Dirac cone (blue). The ${\cal H}_{\rm R}({\bf k})$ is dominant, gapping the top surface Dirac cone (red). 
[(a2)-(c2)] Hall conductance as a function of the Fermi energy $E_{F}$, corresponding to the parameters in (a1) $T_{1}=0.3$ $\hbar/$eV, $T_{2}=0.1$ $\hbar/$eV; (b1) $T_{1}=T_{2}=0.1$ $\hbar/$eV; and (c1) $T_{1}=0.1$ $\hbar/$eV, $T_{2}=0.3$ $\hbar/$eV. They exhibit zero (axion), half-integer (semi-magnetic), and integer-quantized (Chern) Hall conductivity in the gap, respectively. 
Here, $V_z=0.1$ eV and $A_{0}=0.8$ nm$^{-1}$.
The other parameters are the same as those in Fig.~\ref{Fig:E_C_Lz30_delta163_inverse_together}.
} \label{Fig:Quench_E_C_Lz30_togethe_030103}
\end{figure}

An interesting extension of the above-mentioned approach involves performing a Floquet quench on the axion and Chern phases. Since they are generated by left and right polarized light, we shall investigate the possible outcomes of periodically quenching the chirality of the circularly polarized light to alternate rapidly between Figs.~\ref{Fig:Schematic}(e) and ~\ref{Fig:Schematic}(f) [or Figs.~\ref{Fig:Schematic_light}(a) and ~\ref{Fig:Schematic_light}(b)]. Without loss of generality, we take Figs.~\ref{Fig:Schematic}(e) (left-handed circularly polarized light with $\varphi=-\pi/2$) and ~\ref{Fig:Schematic}(f) (right-handed circularly polarized light with $\varphi=\pi/2$) in the following discussions.

We consider a periodic two-step quench with a total period of $T=T_{1}+T_{2}$, where each odd(even) step is governed by the Hamiltonian under left (right) polarized light ${\cal H}_{\rm L}({\bf k})$ [${\cal H}_{\rm R}({\bf k})$], for a duration of $T_{1}$ [$T_{2}$]. Then the effective Floquet Hamiltonian is given by~\cite{xiong2016towards,liu2019floquet,li2018realistic}
\begin{align}
{\cal H}_\text{eff}({\bf k})\equiv\frac{i}{T_{1}+T_{2}}\ln[e^{-i{\cal H}_{\rm R}({\bf k})T_{2}}e^{-i{\cal H}_{\rm L}({\bf k})T_{1}}],\label{eq:Heff}
\end{align} where ${\cal H}_{\rm L}({\bf k})$ denotes the Hamiltonian under left-handed circularly polarized light with $\varphi=-\pi/2$ and ${\cal H}_{\rm R}({\bf k})$ denotes the Hamiltonian under right-handed circularly polarized light with $\varphi=\pi/2$. 

As shown in Fig.~\ref{Fig:Quench_E_C_Lz30_togethe_030103}, we can find that the value of the gap of the top surface (red curve) can be tuned by the time duration parameters $T_{1}$ and $T_{2}$. In particular, when $T_{1}=T_{2}$ as shown in Fig.~\ref{Fig:Quench_E_C_Lz30_togethe_030103}(b1), the system becomes a semi-magnetic topological insulator phase, which is different from the original axion and Chern insulator phases. Finally, we note that higher half-integer quantized conductivities can also be realized through Floquet driving protocols in slightly more complicated related settings~\cite{yap2018photoinduced}.

\section{Summary and Discussion}\label{7}

We propose a 3D topological insulator heterostructure that can exhibit a variety of topologically quantized or half-quantized phases through selective magnetic doping and/or irradiation with circularly polarized lasers, both of which open up a gap in the surface Dirac cones through time-reversal breaking. In particular, when magnetic ions are modulation-doped only in the vicinity of the bottom surface and high-frequency circularly polarized light is irradiated into the top surface (without penetrating the bottom), we can either realize the axion insulator (with zero Hall plateau) or the Chern insulator (with quantized Hall plateau) by toggling the polarization chirality. These results are substantiated by explicit evaluation of the Floquet Hamiltonian and numerical computation of the resultant Berry curvatures based on realistic topological insulator material parameters such as the optical penetration depth.

It is interesting to note that although all the above-mentioned phases are defined in 3D topological systems, they can be mathematically ``compressed'' into 2D time-reversal broken systems via an inverse holographic mapping~\cite{gu2016holographic,qi2013exact,lee2016exact,lee2017generalized}, such that the layer-resolved Berry curvatures become the scale-resolved Berry curvatures of the corresponding 2D holographic duals. With that, the 3D Hall conductivities correspond directly to 2D dual Chern numbers, and half-quantized 3D insulators correspond to non-lattice regularized 2D gapped Dirac cones. In this paper, the main effect of Floquet driving was to gap out Dirac cones through time-reversal breaking; the investigation of how this mechanism interplays with decidedly more robust nonlinear Dirac cones~\cite{bomantara2017nonlinear,tuloup2020nonlinearity,tuloup2022topological} would certainly be interesting for future investigations.

\begin{acknowledgments} 
F.Q. is supported by the QEP2.0 Grant from the Singapore National Research Foundation (Grant No.~NRF2021-QEP2-02-P09) and the MOE Tier-II Grant (Proposal ID: T2EP50222-0008). R.C. acknowledges the support from the Chutian Scholars Program in Hubei Province and the National Natural Science Foundation of China (Grant No. 12304195).
\end{acknowledgments}

\appendix
\onecolumngrid



\section{Derivation for the momentum-space tight-binding model}\label{Appendix_A}

The motivation of this Appendix \ref{Appendix_A} is to derive the analytical expression of the momentum-space tight-binding model for both (Bi,Sb)$_{2}$Te$_{3}$ and Bi$_2$Se$_3$ without Cr doping and optical (Floquet) driving.

We begin from the low-energy three-dimensional effective model Hamiltonian for bulk (Bi,Sb)$_{2}$Te$_{3}$ and Bi$_2$Se$_3$ near the $\Gamma$ point, which is given in the basis $(|1p_{z}^{+},\uparrow\rangle , |2p_{z}^{-},\uparrow\rangle , |1p_{z}^{+},\downarrow\rangle , |2p_{z}^{-},\downarrow\rangle )$ which are the hybridized states of Te or Se $p_{z}$ orbital (1) and Sb or Bi $p_{z}$ orbital (2), with even ($+$) and odd ($-$) parities, up ($\uparrow$) and down ($\downarrow$) spins~\cite{yu2010quantized,lu2010massive,shan2010effective}, as used in~\cite{shan2010effective,lu2010massive,li2010chern,lu2013quantum,chen2019effects,sun2020analytical,liu2010model,dabiri2021light,dabiri2021engineering,qin2022phase}
\begin{align}\label{eq:H0_S}
{\cal H}^{(0)}({\bf k}) 
&\!=\!\begin{pmatrix}
m({\bf k}) & A_{1}k_{z} & 0 & A_{2}k_{-} \\
A_{1}k_{z} & -m({\bf k}) & A_{2}k_{-}  & 0 \\
0 & A_{2}k_{+} & m({\bf k}) & -A_{1}k_{z} \\
A_{2}k_{+} & 0 & -A_{1}k_{z} & -m({\bf k})
\end{pmatrix} \\
&\!=\! m({\bf k})\sigma_{0}\otimes\tau_{z} \!+\! \!A_{1}k_{z}\sigma_{z}\otimes\tau_{x} \!+\! A_{2}k_{x}\sigma_{x}\otimes\tau_{x} \!+\! A_{2}k_{y}\sigma_{y}\otimes\tau_{x},\label{eq:H_3D_S}
\end{align} where $\sigma_{x,y,z}$ are the Pauli matrices for the spin degree of freedom, $\tau_{x,y,z}$ are the Pauli matrices for the orbital degree of freedom, $\sigma_{0}$ ($\tau_{0}$) is a $2\times2$ unit matrix, $m({\bf k})=m_{0}-B_{1}k_{z}^{2}-B_{2}(k_{x}^{2}+k_{y}^{2})$, $k_{\pm}=k_{x}\pm ik_{y}$, $m_0$, $A_1$, $A_2$, $B_1$, and $B_2$ are model parameters to be specified later.

To regularize the low-energy long-wavelength Hamiltonian on a lattice, one makes the following replacements~\cite{shen2017topological}:
\begin{align}
&k_{j}\rightarrow\frac{1}{a_{j}}\sin(k_{j}a_{j}),\label{eq:sin}\\
&k_{j}^{2}\rightarrow\frac{2}{a_{j}^2}[1-\cos(k_{j}a_{j})],\label{eq:cos}
\end{align} where $j=x,y,z$ and $a_j$ is the lattice constant along $j$ direction.
With the mappings \eqref{eq:sin} and \eqref{eq:cos}, one obtains 
\begin{align}
&M({\bf k})=m_{0}-\frac{2B_{1}}{a_{z}^2}[1-\cos(k_{z}a_{z})]-\frac{2B_{2}}{a_{x}^2}[1-\cos(k_{x}a_{x})]-\frac{2B_{2}}{a_{y}^2}[1-\cos(k_{y}a_{y})],\\
&\tilde{k}_{\pm}=\frac{1}{a_{x}}\sin(k_{x}a_{x})\pm \frac{i}{a_{y}}\sin(k_{y}a_{y}).
\end{align}
Due to the lattice symmetry, $a_{x}=a_{y}=a_{||}$ and we have
\begin{align}
M({\bf k})&=m_{0}-\frac{2B_{1}}{a_{z}^2}[1-\cos(k_{z}a_{z})]-\frac{2B_{2}}{a_{||}^2}[2-\cos(k_{x}a_{||})-\cos(k_{y}a_{||})] \nonumber\\
&=m_{0}-\frac{4B_{1}}{a_{z}^2}\sin^{2}\frac{k_{z}a_{z}}{2}-\frac{4B_{2}}{a_{||}^2}\left(\sin^{2}\frac{k_{x}a_{||}}{2}+\sin^{2}\frac{k_{y}a_{||}}{2}\right),\\
\tilde{k}_{\pm}&=\frac{1}{a_{||}}[\sin(k_{x}a_{||})\pm i\sin(k_{y}a_{||})].
\end{align}
Therefore, the momentum-space tight-binding model for both (Bi,Sb)$_{2}$Te$_{3}$ and Bi$_2$Se$_3$ without Cr doping and light in the basis $(c_{{\bf k},+,\uparrow}^{}, c_{{\bf k},-,\uparrow}^{}, c_{{\bf k},+,\downarrow}^{}, c_{{\bf k},-,\downarrow}^{})^{T}$ is given by
\begin{align}
{\cal H}_{\rm 3D}({\bf k})\!=\! M({\bf k})\sigma_{0}\otimes\tau_{z} \!+\! \!\frac{A_{1}}{a_{z}}\sin(k_{z}a_{z})\sigma_{z}\otimes\tau_{x} \!+\! \frac{A_{2}}{a_{||}}\sin(k_{x}a_{||})\sigma_{x}\otimes\tau_{x} \!+\! \frac{A_{2}}{a_{||}}\sin(k_{y}a_{||})\sigma_{y}\otimes\tau_{x}.
\end{align}

By setting that $t_{z}=\frac{B_{1}}{a_{z}^2}$, $t_{||}=\frac{B_{2}}{a_{||}^2}$, $\lambda_{z}=\frac{A_{1}}{a_{z}}$, and $\lambda_{||}=\frac{A_{2}}{a_{||}}$, we have
\begin{align}
{\cal H}_{\rm 3D}({\bf k}) 
&=M({\bf k})\sigma_{0}\otimes\tau_{z}\!+\!\lambda_{z}\sin(k_{z}a_{z})\sigma_{z}\otimes\tau_{x} \!+\! \lambda_{||}\sin(k_{x}a_{||})\sigma_{x}\otimes\tau_{x} \!+\! \lambda_{||}\sin(k_{y}a_{||})\sigma_{y}\otimes\tau_{x},\label{eq:Htb_3D_S}\\
M({\bf k})&=m_{0}-4t_{z}\sin^{2}\frac{k_{z}a_{z}}{2}-4t_{||}\left(\sin^{2}\frac{k_{x}a_{||}}{2}+\sin^{2}\frac{k_{y}a_{||}}{2}\right) \nonumber\\
&=m_{0}-4t_{z}\sin^{2}\frac{k_{z}a_{z}}{2}-2t_{||}\left[1-\cos(k_{x}a_{||})+1-\cos(k_{y}a_{||})\right] \nonumber\\
&=(m_{0}-2t_{z}-4t_{||})+2t_{z}\cos(k_{z}a_{z})+2t_{||}\left[\cos(k_{x}a_{||})+\cos(k_{y}a_{||})\right].\label{eq:M_k_S}
\end{align}
The parameters for both (Bi,Sb)$_{2}$Te$_{3}$ and Bi$_2$Se$_3$ are adopted as~\cite{mogi2022experimental,zou2023half,zhang2009topological,chang2013experimental}: $A_{1}=0.22$ eV$\cdot$nm, $A_{2}=0.41$ eV$\cdot$nm, $m_{0}=0.28$ eV, $B_{1}=0.1$ eV$\cdot$nm$^2$, and $B_{2}=0.566$ eV$\cdot$nm$^2$. Furthermore, by setting $a_{z}=0.5$ nm and $a_{||}=a=1$ nm, we can have $t_{z}=0.4$ eV, $t_{||}=0.566$ eV, $\lambda_{z}=0.44$ eV, and $\lambda_{||}=0.41$ eV.

\section{Derivation for the real-space tight-binding model \eqref{eq:H0}}\label{Appendix_B}

The motivation of this Appendix \ref{Appendix_B} is to derive the analytical expression of the real-space tight-binding model for both (Bi,Sb)$_{2}$Te$_{3}$ and Bi$_2$Se$_3$ without Cr doping and optical driving, under open boundary conditions along the $z$ direction and periodic boundary conditions along the $x$ and $y$ directions. This is necessary for the numerical calculations in the main text.

In momentum space, the low-energy three-dimensional tight-binding model Hamiltonian for topological insulators (Bi,Sb)$_{2}$Te$_{3}$ and Bi$_2$Se$_3$ is given by~\cite{shen2017topological,zou2023half,chu2011surface,zhang2009topological,liu2010model,ding2020hinged} ($a_{x}=a_{y}=a_{||}=a$)
\begin{align}
{\cal H}_{\rm 3D}({\bf k}) 
\!=\!M({\bf k})\sigma_{0}\otimes\tau_{z}\!+\!\lambda_{z}\sin(k_{z}a_{z})\sigma_{z}\otimes\tau_{x} \!+\! \lambda_{||}\sin(k_{x}a)\sigma_{x}\otimes\tau_{x} \!+\! \lambda_{||}\sin(k_{y}a)\sigma_{y}\otimes\tau_{x},
\end{align} where $M({\bf k})$ is given by Eq.~\eqref{eq:M_k_S}, $\sigma_{x,y,z}$ and $\tau_{x,y,z}$ are the Pauli matrices for the spin and orbital degrees of freedom, respectively, $\sigma_{0}$ ($\tau_{0}$) is a $2\times2$ unit matrix, the wave vector is ${\bf k}=(k_x,k_y,k_z)$, $a$ is the lattice constant, $m_0$, $t_{z}$, $t_{||}$, $\lambda_{z}$, and $\lambda_{||}$ are model parameters. The parameters for both (Bi,Sb)$_{2}$Te$_{3}$ and Bi$_2$Se$_3$ are adopted as~\cite{mogi2022experimental,zou2023half}: $t_{z}=0.40$ eV, $t_{||}=0.566$ eV, $\lambda_{z}=0.44$ eV, $\lambda_{||}=0.41$ eV, and $m_{0}=0.28$ eV.

Fourier transforming along the $z$ direction so as to go from momentum to real space, one has~\cite{shen2017topological}
\begin{align}
c_{{\bf k},s,\sigma}^{\dagger}=\frac{1}{\sqrt{N_{z}}}\sum_{j_z=1}^{N_z}e^{ik_{z}j_{z}a_{z}}c_{k_x,k_y,j_z,s,\sigma}^{\dagger},~~~~
c_{{\bf k},s,\sigma}^{}=\frac{1}{\sqrt{N_{z}}}\sum_{j_z=1}^{N_z}e^{-ik_{z}j_{z}a_{z}}c_{k_x,k_y,j_z,s,\sigma}^{},
\end{align} 
\begin{align}
C_{{\bf k}}^{\dagger}=\begin{pmatrix}
c_{{\bf k},+,\uparrow}^{\dagger} & c_{{\bf k},-,\uparrow}^{\dagger} & c_{{\bf k},+,\downarrow}^{\dagger} &
c_{{\bf k},-,\downarrow}^{\dagger}
\end{pmatrix}=\frac{1}{\sqrt{N_{z}}}\sum_{j_z=1}^{N_z}e^{ik_{z}j_{z}a_{z}}C_{k_x,k_y,j_z}^{\dagger},~
C_{{\bf k}}^{}=\begin{pmatrix}
c_{{\bf k},+,\uparrow}^{} \\
c_{{\bf k},-,\uparrow}^{} \\
c_{{\bf k},+,\downarrow}^{} \\
c_{{\bf k},-,\downarrow}^{}
\end{pmatrix}=\frac{1}{\sqrt{N_{z}}}\sum_{j_z=1}^{N_z}e^{-ik_{z}j_{z}a_{z}}C_{k_x,k_y,j_z}^{}.
\end{align}

As such, we obtain
\begin{align}
C_{{\bf k}}^{\dagger}\left[{\cal H}_{\rm 3D}({\bf k})\right]C_{{\bf k}} 
&\!=\!\sum_{j_z}\!\left[\!m_{0}-2t_{z}-4t_{||}\left(\sin^{2}\frac{k_{x}a}{2}+\sin^{2}\frac{k_{y}a}{2}\right)\!\right]\!C_{k_x,k_y,j_z}^{\dagger}[\sigma_{0}\otimes\tau_{z}]C_{k_x,k_y,j_z} \nonumber\\
&\!+\!t_{z}\sum_{j_z}\!\left\{\!C_{k_x,k_y,j_z}^{\dagger}[\sigma_{0}\otimes\tau_{z}]C_{k_x,k_y,j_z+1} + C_{k_x,k_y,j_z+1}^{\dagger}[\sigma_{0}\otimes\tau_{z}]C_{k_x,k_y,j_z} \!\right\}\! \nonumber\\ 
&\!-\! i\frac{\lambda_{z}}{2}\sum_{j_z}\!\left\{\!C_{k_x,k_y,j_z}^{\dagger}[\sigma_{z}\otimes\tau_{x}]C_{k_x,k_y,j_z+1}\!-\! C_{k_x,k_y,j_z+1}^{\dagger}[\sigma_{z}\otimes\tau_{x}]C_{k_x,k_y,j_z}\!\right\}\! \nonumber\\
& \!+\! \lambda_{||}\sin(k_{x}a)\sum_{j_z}C_{k_x,k_y,j_z}^{\dagger}[\sigma_{x}\otimes\tau_{x}]C_{k_x,k_y,j_z} \!+\! \lambda_{||}\sin(k_{y}a)\sum_{j_z}C_{k_x,k_y,j_z}^{\dagger}[\sigma_{y}\otimes\tau_{x}]C_{k_x,k_y,j_z},
\end{align} where we have used $\sin^{2}\frac{k_{z}a_{z}}{2}=\left(\frac{e^{ik_{z}a_{z}/2}-e^{-ik_{z}a_{z}/2}}{2i}\right)^{2}=-\frac{1}{4}(e^{ik_{z}a_{z}}+e^{-ik_{z}a_{z}}-2)$.

Therefore, the real-space tight-binding Hamiltonian under $x$-PBCs, $y$-PBCs, and $z$-OBCs in the basis $(C_{k_x,k_y,1}^{}, C_{k_x,k_y,2}^{},\cdots,C_{k_x,k_y,N_z}^{})^{T}$ is given by 
\begin{align}
{\cal H}_{\rm tb}^{(0)}=\begin{pmatrix}
h & T_{z} & 0 & \cdots & 0 \\
T_{z}^{\dagger} & h & T_{z} & \cdots & 0 \\
0 & T_{z}^{\dagger} & h & \ddots & \vdots \\
\vdots& \ddots & \ddots & \ddots & T_{z} \\
0 & \cdots & 0 & T_{z}^{\dagger} & h
\end{pmatrix}_{4N_z\times 4N_z},
\end{align} where
\begin{align}
h&\!=\!\left[m_{0}-2t_{z}-4t_{||}\left(\sin^{2}\frac{k_{x}a}{2}+\sin^{2}\frac{k_{y}a}{2}\right) \right]\sigma_{0}\otimes\tau_{z} \!+\! \lambda_{||}\sin(k_{x}a)\sigma_{x}\otimes\tau_{x} \!+\! \lambda_{||}\sin(k_{y}a)\sigma_{y}\otimes\tau_{x},\\
&\!=\!\left\{m_{0}-2t_{z}-4t_{||}+2t_{||}\left[\cos(k_{x}a)+\cos(k_{y}a)\right] \right\}\sigma_{0}\otimes\tau_{z} \!+\! \lambda_{||}\sin(k_{x}a)\sigma_{x}\otimes\tau_{x} \!+\! \lambda_{||}\sin(k_{y}a)\sigma_{y}\otimes\tau_{x}\\
&\!=\!\left[(m_{0}-2t_{z}-4t_{||})+t_{||}\left(e^{ik_{x}a}+e^{-ik_{x}a}+e^{ik_{y}a}+e^{-ik_{y}a}\right) \right]\sigma_{0}\otimes\tau_{z} \nonumber\\
&~~~\!-\! i\frac{\lambda_{||}}{2}\left(e^{ik_{x}a}-e^{-ik_{x}a}\right)\sigma_{x}\otimes\tau_{x}\!-\! i\frac{\lambda_{||}}{2}\left(e^{ik_{y}a}-e^{-ik_{y}a}\right)\sigma_{y}\otimes\tau_{x} \nonumber\\
&\!=\!M_{0}+T_{x}e^{ik_{x}a}+T_{x}^{\dagger}e^{-ik_{x}a}+T_{y}e^{ik_{y}a}+T_{y}^{\dagger}e^{-ik_{y}a},\\
M_{0}&\!=\!\left(m_{0}-2t_{z}-4t_{||}\right)\sigma_{0}\otimes\tau_{z},\\
T_{x}&\!=\!t_{||}\sigma_{0}\otimes\tau_{z}\!-\! i\frac{\lambda_{||}}{2}\sigma_{x}\otimes\tau_{x},~~
T_{x}^{\dagger}\!=\!t_{||}\sigma_{0}\otimes\tau_{z}\!+\! i\frac{\lambda_{||}}{2}\sigma_{x}\otimes\tau_{x},\\
T_{y}&\!=\!t_{||}\sigma_{0}\otimes\tau_{z}\!-\! i\frac{\lambda_{||}}{2}\sigma_{y}\otimes\tau_{x},~~
T_{y}^{\dagger}\!=\!t_{||}\sigma_{0}\otimes\tau_{z}\!+\! i\frac{\lambda_{||}}{2}\sigma_{y}\otimes\tau_{x},\\
T_{z}&\!=\!t_{z}\sigma_{0}\otimes\tau_{z}\!-\! i\frac{\lambda_{z}}{2}\sigma_{z}\otimes\tau_{x},~~
T_{z}^{\dagger}\!=\!t_{z}\sigma_{0}\otimes\tau_{z}\!+\! i\frac{\lambda_{z}}{2}\sigma_{z}\otimes\tau_{x}.
\end{align}

\section{The matrix form of the real-space tight-binding model with Cr doping}\label{Appendix_C}

Here, we derive the matrix form of the real-space tight-binding model with Cr doping under open boundary conditions along the $z$ direction and periodic boundary conditions along the $x$ and $y$ directions.

The real-space tight-binding Hamiltonian with Cr doping without light under $x$-PBCs, $y$-PBCs, and $z$-OBCs in the basis $(C_{k_x,k_y,1}^{}, C_{k_x,k_y,2}^{},\cdots,C_{k_x,k_y,N_z}^{})^{T}$ is given by 
\begin{align}
{\cal H}_{\rm tb}={\cal H}_{\rm tb}^{(0)} + \Delta {\cal H}_{ex},
\end{align} where
\begin{align}
\Delta {\cal H}_{ex}=\begin{pmatrix}
\Delta{\cal H}_{d}(1) & 0 & 0 & \cdots & 0 \\
0 & \Delta{\cal H}_{d}(2) & 0 & \cdots & 0 \\
0 & 0 & \Delta{\cal H}_{d}(3) & \ddots & \vdots \\
\vdots& \ddots & \ddots & \ddots & 0 \\
0 & \cdots & 0 & 0 & \Delta{\cal H}_{d}(N_z)
\end{pmatrix}_{4N_z\times 4N_z},
\end{align} and 
\begin{align}
\Delta{\cal H}_{d}(j_z)\!=\!V_{z}(j_z)\sigma_{z}\otimes\tau_{0}.
\end{align}

\section{Time-reversal symmetry breaking with magnetic doping}\label{Appendix_D}

Here we show that when the magnetic doping is added, i.e., $V_{z}(j_z)\neq 0$, time-reversal symmetry is broken.

The Hamiltonian with magnetic doping under time-reversal transformation becomes
\begin{align}\label{eq:time_reversal_m0}
{\cal T}[H({\bf k})]{\cal T}^{-1}
&\!=\!\sum_{j_{z}}C^{\dagger}_{{\bf k},j_{z}}\sigma_{0}\otimes\tau_{y}{\cal K}\left\{ \left[m_{0}-2t_{z}-4t_{||}\left(\sin^{2}\frac{k_{x}a}{2}+\sin^{2}\frac{k_{y}a}{2}\right) \right]\sigma_{0}\otimes\tau_{z} \right.\nonumber\\
&\left. \!+\! \lambda_{||}\sin(k_{x}a)\sigma_{x}\otimes\tau_{x} \!+\! \lambda_{||}\sin(k_{y}a)\sigma_{y}\otimes\tau_{x} \right\}{\cal K}^{-1}(\sigma_{0}\otimes\tau_{y})^{-1}C_{{\bf k},j_{z}} \nonumber\\
&\!+\!\sum_{j_{z}}\left[C^{\dagger}_{{\bf k},j_{z}}\sigma_{0}\otimes\tau_{y}{\cal K}T{\cal K}^{-1}(\sigma_{0}\otimes\tau_{y})^{-1}C_{{\bf k},j_{z}+1} + C^{\dagger}_{{\bf k},j_{z}+1}\sigma_{0}\otimes\tau_{y}{\cal K}T^{\dagger}{\cal K}^{-1}(\sigma_{0}\otimes\tau_{y})^{-1}C_{{\bf k},j_{z}} \right] \nonumber\\
&\!+\!\sum_{j_{z}}C^{\dagger}_{{\bf k},j_{z}}V_{z}(j_z)\sigma_{0}\otimes\tau_{y}{\cal K}\left( \sigma_{z}\otimes\tau_{0} \right){\cal K}^{-1}(\sigma_{0}\otimes\tau_{y})^{-1}C_{{\bf k},j_{z}} \nonumber\\
&\!=\!\sum_{j_{z}}C^{\dagger}_{{\bf k},j_{z}}\left\{ \left[m_{0}-4t_{z}\sin^{2}\frac{k_{z}a}{2}-4t_{||}\left(\sin^{2}\frac{k_{x}a}{2}+\sin^{2}\frac{k_{y}a}{2}\right)\right]\sigma_{0}\otimes\tau_{z} \right.\nonumber\\
&\left. \!-\! \lambda_{||}\sin(k_{x}a)\sigma_{x}\otimes\tau_{x} \!-\! \lambda_{||}\sin(k_{y}a)\sigma_{y}\otimes\tau_{x} \!-\!V_{z}(j_z)\sigma_{z}\otimes\tau_{0} \right\}C_{{\bf k},j_{z}} \nonumber\\
&\!+\!\sum_{j_{z}}\left[C^{\dagger}_{{\bf k},j_{z}}TC_{{\bf k},j_{z}+1} + C^{\dagger}_{{\bf k},j_{z}+1}T^{\dagger}C_{{\bf k},j_{z}} \right] \nonumber\\
&\!=\! H(-{\bf k})\!-\!\sum_{j_{z}}C^{\dagger}_{{\bf k},j_{z}}2V_{z}(j_z)\sigma_{z}\otimes\tau_{0}C_{{\bf k},j_{z}},
\end{align}
where $H({\bf k})=H^{(0)}({\bf k})+\Delta H_{d}$, $H(-{\bf k})=H^{(0)}(-{\bf k})+\Delta H_{d}$, ${\cal T}=\sigma_{y}\otimes\tau_{0}{\cal K}$~\cite{schindler2018higher} is the time-reversal operator with the complex conjugate operator ${\cal K}$, we have ${\cal K}H({\bf k}){\cal K}^{-1}=H^{*}({\bf k})$, and we use
\begin{align}
&(\sigma_{y}\otimes\tau_{0})(\sigma_{0}\otimes\tau_{z})(\sigma_{y}\otimes\tau_{0})^{-1}=\sigma_{0}\otimes\tau_{z} ,\\
&(\sigma_{y}\otimes\tau_{0})(\sigma_{z}\otimes\tau_{x})(\sigma_{y}\otimes\tau_{0})^{-1}=-\sigma_{z}\otimes\tau_{x} ,\\
&(\sigma_{y}\otimes\tau_{0})(\sigma_{x}\otimes\tau_{x})(\sigma_{y}\otimes\tau_{0})^{-1}=-\sigma_{x}\otimes\tau_{x} ,\\
&(\sigma_{y}\otimes\tau_{0})({\cal K}\sigma_{y}\otimes\tau_{x}{\cal K}^{-1})(\sigma_{y}\otimes\tau_{0})^{-1}=-\sigma_{y}\otimes\tau_{x} ,\\
&(\sigma_{y}\otimes\tau_{0})({\cal K}T{\cal K}^{-1})(\sigma_{y}\otimes\tau_{0})^{-1}=T,\\
&(\sigma_{y}\otimes\tau_{0})({\cal K}T^{\dagger}{\cal K}^{-1})(\sigma_{y}\otimes\tau_{0})^{-1}=T^{\dagger},\\
&(\sigma_{y}\otimes\tau_{0})(\sigma_{z}\otimes\tau_{0})(\sigma_{y}\otimes\tau_{0})^{-1}=-\sigma_{z}\otimes\tau_{0}.
\end{align}
As a result of Eq.~(\ref{eq:time_reversal_m0}), if $V_{z}(j_z)=0$, we have ${\cal T}H({\bf k}){\cal T}^{-1}=H(-{\bf k})$ which shows a time-reversal symmetry.
However, for $V_{z}(j_z)\neq0$, Eq.~(\ref{eq:time_reversal_m0}) shows that the time-reversal symmetry is broken.

\section{Expressions for $H_{0}$, $H_{-n}$, and $H_{n}$}\label{Appendix_E}

The motivation for this Appendix \ref{Appendix_E} is to derive the concrete analytical expressions for the time Fourier components $H_{0}$, $H_{-n}$, and $H_{n}$ that enter the Floquet Hamiltonian (\ref{eq:HF0}) of the main text.

With ${\bf A}(z,t)=\omega^{-1}E_{0}e^{-z/\delta}( \sin(\omega t),  \sin(\omega t + \varphi), 0 )$, $A(z)=A_{0}e^{-z/\delta}$ and $A_{0}=eE_{0}/(\hbar\omega)$, we have the photon-dressed effective Hamiltonian as 
\begin{align}
H^{(0)}({\bf k},t)&\!=\! \sum_{j_{z}}C^{\dagger}_{{\bf k},j_{z}}\left\{m_{0}-2t_{z}-4t_{||}\left[\sin^{2}\frac{\left(k_{x}-A(z)\sin(\omega t)\right)a}{2}+\sin^{2}\frac{\left(k_{y}-A(z)\sin(\omega t + \varphi)\right)a}{2}\right]\right\}\sigma_{0}\otimes\tau_{z}C_{{\bf k},j_{z}} \nonumber\\
&~~\!+\! \sum_{j_{z}}C^{\dagger}_{{\bf k},j_{z}}\left\{\lambda_{||}\sin[(k_{x}-A(z)\sin(\omega t))a]\sigma_{x}\otimes\tau_{x} \!+\! \lambda_{||}\sin[(k_{y}-A(z)\sin(\omega t + \varphi))a]\sigma_{y}\otimes\tau_{x}\right\}C_{{\bf k},j_{z}} \nonumber\\
&~~\!+\!\sum_{j_{z}}\left(C^{\dagger}_{{\bf k},j_{z}}T_{z}C_{{\bf k},j_{z}+1} + C^{\dagger}_{{\bf k},j_{z}+1}T_{z}^{\dagger}C_{{\bf k},j_{z}} \right) \nonumber\\
&\!=\!\sum_{j_{z}}C^{\dagger}_{{\bf k},j_{z}}\left\{m_{0}-2t_{z}-4t_{||} \right.\nonumber\\
&\left.~~+t_{||}\left[e^{i[k_{x}-A(z)\sin(\omega t)]a}+e^{-i[k_{x}-A(z)\sin(\omega t)]a}+e^{i[k_{y}-A(z)\sin(\omega t + \varphi)]a}+e^{-i[k_{y}-A(z)\sin(\omega t + \varphi)]a}\right]\right\}\sigma_{0}\otimes\tau_{z}C_{{\bf k},j_{z}} \nonumber\\
&~~\!+\! \sum_{j_{z}}C^{\dagger}_{{\bf k},j_{z}}\frac{\lambda_{||}}{2i}\left[e^{i[k_{x}-A(z)\sin(\omega t)]a}-e^{-i[k_{x}-A(z)\sin(\omega t)]a}\right]\sigma_{x}\otimes\tau_{x}C_{{\bf k},j_{z}} \nonumber\\
&~~\!+\! \sum_{j_{z}}C^{\dagger}_{{\bf k},j_{z}}\frac{\lambda_{||}}{2i}\left[e^{i[k_{y}-A(z)\sin(\omega t + \varphi)]a}-e^{-i[k_{y}-A(z)\sin(\omega t + \varphi)]a}\right]\sigma_{y}\otimes\tau_{x}C_{{\bf k},j_{z}} \nonumber\\
&~~\!+\!\sum_{j_{z}}\left(C^{\dagger}_{{\bf k},j_{z}}T_{z}C_{{\bf k},j_{z}+1} + C^{\dagger}_{{\bf k},j_{z}+1}T_{z}^{\dagger}C_{{\bf k},j_{z}} \right).
\end{align}
Furthermore, the concrete analytical expressions for $H_{0}$, $H_{-n}$, and $H_{n}$ in the Floquet Hamiltonian (\ref{eq:HF0}) are given as 
\begin{align}
H_{0}&\!=\! \frac{1}{T} \int_{0}^{T}H^{(0)}({\bf k},t) dt \nonumber\\
&\!=\! \sum_{j_{z}}C^{\dagger}_{{\bf k},j_{z}}\left\{\left\{m_{0}-2t_{z}-4t_{||}+2{\cal J}_{0}(A(z)a)t_{||}\left[\cos(k_{x}a)+\cos(k_{y}a)\right]\right\}\sigma_{0}\otimes\tau_{z} \right.\nonumber\\
&\left.~~\!+\! {\cal J}_{0}(A(z)a)\lambda_{||}\sin(k_{x}a)\sigma_{x}\otimes\tau_{x}\!+\! {\cal J}_{0}(A(z)a)\lambda_{||}\sin(k_{y}a)\sigma_{y}\otimes\tau_{x}\right\}C_{{\bf k},j_{z}} \nonumber\\
&~~\!+\!\sum_{j_{z}}\left(C^{\dagger}_{{\bf k},j_{z}}T_{z}C_{{\bf k},j_{z}+1} + C^{\dagger}_{{\bf k},j_{z}+1}T_{z}^{\dagger}C_{{\bf k},j_{z}} \right),\label{eq:H00}\\
\!H_{-n}&\!=\! \frac{1}{T} \int_{0}^{T}H^{(0)}({\bf k},t) e^{-in\omega t}dt \nonumber\\
&\!=\! \sum_{j_{z}}C^{\dagger}_{{\bf k},j_{z}}\left\{{\cal J}_{n}(A(z)a)t_{||}\left\{(-1)^{n}e^{ik_{x}a}+e^{-ik_{x}a}+e^{in\varphi}\left[(-1)^{n}e^{ik_{y}a}+e^{-ik_{y}a}\right]\right\}\sigma_{0}\otimes\tau_{z} \right.\nonumber\\
&\left.~~\!+\! \frac{{\cal J}_{n}(A(z)a)\lambda_{||}}{2i}\left[(-1)^{n}e^{ik_{x}a}-e^{-ik_{x}a}\right]\sigma_{x}\otimes\tau_{x} 
\!+\! \frac{{\cal J}_{n}(A(z)a)\lambda_{||}}{2i}e^{in\varphi}\left[(-1)^{n}e^{ik_{y}a}-e^{-ik_{y}a}\right]\sigma_{y}\otimes\tau_{x} \right\}C_{{\bf k},j_{z}}, \label{eq:H0-1}\\
\!H_{n}&\!=\! \frac{1}{T} \int_{0}^{T}H^{(0)}({\bf k},t) e^{in\omega t}dt \nonumber\\
&\!=\! \sum_{j_{z}}C^{\dagger}_{{\bf k},j_{z}}\left\{{\cal J}_{n}(A(z)a)t_{||}\left\{e^{ik_{x}a}+(-1)^{n}e^{-ik_{x}a}+e^{-in\varphi}\left[e^{ik_{y}a}+(-1)^{n}e^{-ik_{y}a}\right]\right\}\sigma_{0}\otimes\tau_{z} \right.\nonumber\\
&\left.~~\!+\! \frac{{\cal J}_{n}(A(z)a)\lambda_{||}}{2i}\left[e^{ik_{x}a}-(-1)^{n}e^{-ik_{x}a}\right]\sigma_{x}\otimes\tau_{x} 
\!+\! \frac{{\cal J}_{n}(A(z)a)\lambda_{||}}{2i}e^{-in\varphi}\left[e^{ik_{y}a}-(-1)^{n}e^{-ik_{y}a}\right]\sigma_{y}\otimes\tau_{x} \right\}C_{{\bf k},j_{z}},\label{eq:H01}
\end{align}
where ${\cal J}_{n}(x)=\frac{1}{2\pi}\int_{-\pi}^{\pi}e^{i(n\tau-x\sin\tau)}d\tau$ is the $n$th Bessel function of the first kind~\cite{temme1996special}.

For $n\in{\rm even}$, $n=2n_{0}>0$ with $n_{0}\in\mathbb{N}$ (integers), and $\varphi=\pi/2$, i.e., $e^{\pm in\varphi}=e^{\pm in_{0}\pi}=e^{in_{0}\pi}=\pm1$, we have $H_{-n}=H_{n}$, i.e., $[H_{-n}, H_{n}]=0$.

For $n\in{\rm odd}$, $n=2n_{0}+1>0$ with $n_{0}\in\mathbb{N}$ (integers), and $\varphi=\pi/2$, i.e., $\cos(n\varphi)=\cos\left(\frac{(2n_{0}+1)\pi}{2}\right)=0$, $e^{\pm in\varphi}=\pm i\sin\left(\frac{(2n_{0}+1)\pi}{2}\right)$, we have
\begin{align}
H_{-n}
&\!=\! \sum_{j_{z}}C^{\dagger}_{{\bf k},j_{z}}\left\{-2i{\cal J}_{n}(A(z)a)t_{||}\left[\sin(k_{x}a)+e^{in\varphi}\sin(k_{y}a)\right]\sigma_{0}\otimes\tau_{z} \right.\nonumber\\
&\left.~~\!+\! i{\cal J}_{n}(A(z)a)\lambda_{||}\cos(k_{x}a)\sigma_{x}\otimes\tau_{x} 
\!+\! i{\cal J}_{n}(A(z)a)\lambda_{||}e^{in\varphi}\cos(k_{y}a)\sigma_{y}\otimes\tau_{x} \right\}C_{{\bf k},j_{z}},\label{eq:H0-1_2}\\
H_{n}
&\!=\! \sum_{j_{z}}C^{\dagger}_{{\bf k},j_{z}}\left\{2i{\cal J}_{n}(A(z)a)t_{||}\left[\sin(k_{x}a)+e^{-in\varphi}\sin(k_{y}a)\right]\sigma_{0}\otimes\tau_{z} \right.\nonumber\\
&\left.~~\!-\! i{\cal J}_{n}(A(z)a)\lambda_{||}\cos(k_{x}a)\sigma_{x}\otimes\tau_{x} 
\!-\! i{\cal J}_{n}(A(z)a)\lambda_{||}e^{-in\varphi}\cos(k_{y}a)\sigma_{y}\otimes\tau_{x} \right\}C_{{\bf k},j_{z}},\label{eq:H01_2}
\end{align} 
\begin{align}
&\sum_{n}\frac{[H_{-n}, H_{n}]}{n\hbar\omega} \nonumber\\
&= \sum_{j_{z}}C^{\dagger}_{{\bf k},j_{z}}\sum_{n\in{\rm odd},n>0}\frac{1}{n\hbar\omega}
\left\{ \left(-2i{\cal J}_{n}(A(z)a)t_{||}\left[\sin(k_{x}a)+e^{in\varphi}\sin(k_{y}a)\right] \right)[- i{\cal J}_{n}(A(z)a)\lambda_{||}\cos(k_{x}a)][\gamma_{0}, \gamma_{x}] \right.\nonumber\\
&\left. + \left(2i{\cal J}_{n}(A(z)a)t_{||}\left[\sin(k_{x}a)+e^{-in\varphi}\sin(k_{y}a)\right] \right)[i{\cal J}_{n}(A(z)a)\lambda_{||}\cos(k_{x}a)][\gamma_{x}, \gamma_{0}] \right.\nonumber\\
&\left. + \left(-2i{\cal J}_{n}(A(z)a)t_{||}\left[\sin(k_{x}a)+e^{in\varphi}\sin(k_{y}a)\right] \right)[- i{\cal J}_{n}(A(z)a)\lambda_{||}e^{-in\varphi}\cos(k_{y}a)][\gamma_{0}, \gamma_{y}] \right.\nonumber\\
&\left. + \left(2i{\cal J}_{n}(A(z)a)t_{||}\left[\sin(k_{x}a)+e^{-in\varphi}\sin(k_{y}a)\right] \right)[i{\cal J}_{n}(A(z)a)\lambda_{||}e^{in\varphi}\cos(k_{y}a)][\gamma_{y}, \gamma_{0}] \right.\nonumber\\
&\left. + [i{\cal J}_{n}(A(z)a)\lambda_{||}\cos(k_{x}a)][- i{\cal J}_{n}(A(z)a)\lambda_{||}e^{-in\varphi}\cos(k_{y}a)][\gamma_{x}, \gamma_{y}] \right.\nonumber\\
&\left. + [i{\cal J}_{n}(A(z)a)\lambda_{||}e^{in\varphi}\cos(k_{y}a)][- i{\cal J}_{n}(A(z)a)\lambda_{||}\cos(k_{x}a)][\gamma_{y}, \gamma_{x}] \right\}C_{{\bf k},j_{z}} \nonumber\\
&= \sum_{j_{z}}C^{\dagger}_{{\bf k},j_{z}}\sum_{n\in{\rm odd},n>0}\frac{{\cal J}_{n}^{2}(A(z)a)}{n\hbar\omega}
\left\{ -2t_{||}\lambda_{||}\cos(k_{x}a)\left[\sin(k_{x}a)+e^{in\varphi}\sin(k_{y}a)\right][\gamma_{0}, \gamma_{x}] \right.\nonumber\\
&\left. -2t_{||}\lambda_{||}\cos(k_{x}a)\left[\sin(k_{x}a)+e^{-in\varphi}\sin(k_{y}a)\right][\gamma_{x}, \gamma_{0}] \right.\nonumber\\
&\left. -2t_{||}\lambda_{||}\cos(k_{y}a)\left[e^{-in\varphi}\sin(k_{x}a)+\sin(k_{y}a)\right])[\gamma_{0}, \gamma_{y}] - 2t_{||}\lambda_{||}\cos(k_{y}a)\left[e^{in\varphi}\sin(k_{x}a)+\sin(k_{y}a)\right][\gamma_{y}, \gamma_{0}] \right.\nonumber\\
&\left. + \lambda_{||}^{2}e^{-in\varphi}\cos(k_{y}a)\cos(k_{x}a)[\gamma_{x}, \gamma_{y}] + \lambda_{||}^{2}e^{in\varphi}\cos(k_{y}a)\cos(k_{x}a)[\gamma_{y}, \gamma_{x}] \right\}C_{{\bf k},j_{z}} \nonumber\\
&= \sum_{j_{z}}C^{\dagger}_{{\bf k},j_{z}}\sum_{n\in{\rm odd},n>0}\frac{\lambda_{||}{\cal J}_{n}^{2}(A(z)a)}{n\hbar\omega}
\left\{ -2t_{||}\cos(k_{x}a)\sin(k_{y}a)\left(e^{in\varphi}-e^{-in\varphi}\right)[\gamma_{0}, \gamma_{x}] \right.\nonumber\\
&\left. -2t_{||}\cos(k_{y}a)\sin(k_{x}a)\left(e^{-in\varphi}-e^{in\varphi}\right)[\gamma_{0}, \gamma_{y}] + \lambda_{||}\cos(k_{y}a)\cos(k_{x}a)\left(e^{-in\varphi}-e^{in\varphi}\right)[\gamma_{x}, \gamma_{y}] \right\}C_{{\bf k},j_{z}} \nonumber\\
&= \sum_{j_{z}}C^{\dagger}_{{\bf k},j_{z}}\sum_{n\in{\rm odd},n>0}\frac{\lambda_{||}{\cal J}_{n}^{2}(A(z)a)}{n\hbar\omega}\left(e^{in\varphi}-e^{-in\varphi}\right)
\left\{ -2t_{||}\cos(k_{x}a)\sin(k_{y}a)[\gamma_{0}, \gamma_{x}] 
+2t_{||}\cos(k_{y}a)\sin(k_{x}a)[\gamma_{0}, \gamma_{y}] \right.\nonumber\\
&\left. - \lambda_{||}\cos(k_{y}a)\cos(k_{x}a)[\gamma_{x}, \gamma_{y}] \right\}C_{{\bf k},j_{z}} \nonumber\\
&= \sum_{j_{z}}C^{\dagger}_{{\bf k},j_{z}}\sum_{n\in{\rm odd},n>0}\frac{2i\lambda_{||}{\cal J}_{n}^{2}(A(z)a)}{n\hbar\omega}\sin(n\varphi)
\left\{ 2t_{||}\cos(k_{x}a)\sin(k_{y}a)[\gamma_{x}, \gamma_{0}] 
+2t_{||}\sin(k_{x}a)\cos(k_{y}a)[\gamma_{0}, \gamma_{y}] \right.\nonumber\\
&\left. - \lambda_{||}\cos(k_{x}a)\cos(k_{y}a)[\gamma_{x}, \gamma_{y}] \right\}C_{{\bf k},j_{z}} \\
&= \sum_{j_{z}}C^{\dagger}_{{\bf k},j_{z}}\sum_{n\in{\rm odd},n>0}\frac{2i\lambda_{||}{\cal J}_{n}^{2}(A(z)a)}{n\hbar\omega}\sin\left(\frac{n\pi}{2}\right)
\left\{ 2t_{||}\cos(k_{x}a)\sin(k_{y}a)[\gamma_{x}, \gamma_{0}] 
+2t_{||}\sin(k_{x}a)\cos(k_{y}a)[\gamma_{0}, \gamma_{y}] \right.\nonumber\\
&\left. - \lambda_{||}\cos(k_{x}a)\cos(k_{y}a)[\gamma_{x}, \gamma_{y}] \right\}C_{{\bf k},j_{z}},\label{eq:Hn-n}
\end{align}
where $\gamma_{0}=\sigma_{0}\otimes\tau_{z}$ and $\gamma_{j=x,y,z}=\sigma_{j=x,y,z}\otimes\tau_{x}$.
In numerical calculations, we can choose an appropriate upper cutoff for $n$ by checking whether the results converge consistently independently of $n$.

\section{Matrix form of the real-space tight-binding Floquet Hamiltonian}\label{Appendix_F}

The motivation of this Appendix \ref{Appendix_F} is to derive the matrix form of the real-space tight-binding Floquet Hamiltonian.

Combining Eqs.~\eqref{eq:H00} and \eqref{eq:Hn-n} into \eqref{eq:HF0} of the main text, we have
\begin{align}
H^{(F)}({\bf k}) 
&\!=\!\sum_{j_z}\!\left\{\!m_{0}-2t_{z}-4t_{||}+2{\cal J}_{0}(A(z)a)t_{||}\left[\cos(k_{x}a)+\cos(k_{y}a)\right]\!\right\}\!C_{{\bf k},j_z}^{\dagger}\gamma_{0}C_{{\bf k},j_z} \nonumber\\
&\!+\!t_{z}\sum_{j_z}\!\left(\!C_{{\bf k},j_z}^{\dagger}\gamma_{0}C_{{\bf k},j_z+1} + C_{{\bf k},j_z+1}^{\dagger}\gamma_{0}C_{{\bf k},j_z} \!\right)\! 
\!-\! i\frac{\lambda_{z}}{2}\sum_{j_z}\!\left(\!C_{{\bf k},j_z}^{\dagger}\gamma_{z}C_{{\bf k},j_z+1}\!-\! C_{{\bf k},j_z+1}^{\dagger}\gamma_{z}C_{{\bf k},j_z}\!\right)\! \nonumber\\
& \!+\! {\cal J}_{0}(A(z)a)\lambda_{||}\sin(k_{x}a)\sum_{j_z}C_{{\bf k},j_z}^{\dagger}\gamma_{x}C_{{\bf k},j_z} \!+\! {\cal J}_{0}(A(z)a)\lambda_{||}\sin(k_{y}a)\sum_{j_z}C_{{\bf k},j_z}^{\dagger}\gamma_{y}C_{{\bf k},j_z} \nonumber\\
&\!+\!\sum_{j_z}C_{{\bf k},j_z}^{\dagger}\sum_{n\in{\rm odd},n>0}\frac{2i\lambda_{||}{\cal J}_{n}^{2}(A(z)a)}{n\hbar\omega}\sin\left(\frac{n\pi}{2}\right)
\left\{ 2t_{||}\cos(k_{x}a)\sin(k_{y}a)[\gamma_{x}, \gamma_{0}] 
+2t_{||}\sin(k_{x}a)\cos(k_{y}a)[\gamma_{0}, \gamma_{y}] \right.\nonumber\\
&\left. - \lambda_{||}\cos(k_{x}a)\cos(k_{y}a)[\gamma_{x}, \gamma_{y}] \right\}C_{{\bf k},j_z}.
\end{align}
Therefore, the real-space tight-binding Hamiltonian with light without Cr doping under $x$-PBCs, $y$-PBCs, and $z$-OBCs in the basis $(C_{{\bf k},1}^{}, C_{{\bf k},2}^{},\cdots,C_{{\bf k},N_z}^{})^{T}$ is given by 
\begin{align}
{\cal H}_{F}=\begin{pmatrix}
h_{F} & T_{z} & 0 & \cdots & 0 \\
T_{z}^{\dagger} & h_{F} & T_{z} & \cdots & 0 \\
0 & T_{z}^{\dagger} & h_{F} & \ddots & \vdots \\
\vdots& \ddots & \ddots & \ddots & T_{z} \\
0 & \cdots & 0 & T_{z}^{\dagger} & h_{F}
\end{pmatrix}_{4N_z\times 4N_z},
\end{align} where
\begin{align}
h_{F}&\!=\! \left\{\!m_{0}-2t_{z}-4t_{||}+2{\cal J}_{0}(A(z)a)t_{||}\left[\cos(k_{x}a)+\cos(k_{y}a)\right]\!\right\}\gamma_{0} \!+\! {\cal J}_{0}(A(z)a)\lambda_{||}\sin(k_{x}a)\gamma_{x} \!+\! {\cal J}_{0}(A(z)a)\lambda_{||}\sin(k_{y}a)\gamma_{y} \nonumber\\
&\!+\!\sum_{n\in{\rm odd},n>0}\frac{2i\lambda_{||}{\cal J}_{n}^{2}(A(z)a)}{n\hbar\omega}\sin\left(\frac{n\pi}{2}\right)
\left\{ 2t_{||}\cos(k_{x}a)\sin(k_{y}a)[\gamma_{x}, \gamma_{0}] 
+2t_{||}\sin(k_{x}a)\cos(k_{y}a)[\gamma_{0}, \gamma_{y}] \right.\nonumber\\
&\left. - \lambda_{||}\cos(k_{x}a)\cos(k_{y}a)[\gamma_{x}, \gamma_{y}] \right\},\\
T_{z}&\!=\!t_{z}\gamma_{0}\!-\! i\frac{\lambda_{z}}{2}\gamma_{z},~~
T_{z}^{\dagger}\!=\!t_{z}\gamma_{0}\!+\! i\frac{\lambda_{z}}{2}\gamma_{z}.
\end{align}

\section{Validity of the high-frequency expansion}\label{Appendix_G}

To estimate the validity of the high-frequency expansion quantitatively, we evaluate the maximum instantaneous energy of the time-dependent Hamiltonian $H^{(0)}\left({\bf k} - \frac{e}{\hbar}{\bf A}(z,t)\right)$ averaged over a period of the field $\frac{1}{T}\int_{0}^{T}dt~\text{max}\left\{\big|\big|H^{(0)}({\bf k},t)\big|\big|\right\}<\hbar\omega$ at the $\Gamma$ point ($k_x=k_y= 0$). The optical field parameters have to satisfy the condition $t_{||}(Aa)^{2}/(\hbar\omega)< 1$. In the high-frequency regime $\omega \sim 5.80\times10^{3}$ THz ($\hbar\omega=3.82$ eV)~\cite{humlivcek2014raman}, one can obtain $A(z)=A_{0}e^{-z/\delta}<\frac{1}{a}\sqrt{\frac{\hbar\omega}{t_{||}}}\sim2.59791$ nm$^{-1}$ with $a=1$ nm, i.e., $A_{0}<2.59791e^{z/\delta}$ nm$^{-1}$. With $z\geq0$ and $\delta>0$, one can obtain $A_{0}=eE_{0}/(\hbar\omega)<2.59791$ nm$^{-1}$ ($E_{0}=A_{0}\hbar\omega/e<9.92398\times10^{9}$ V/m), which corresponds to an incident light intensity~\cite{paschotta2016encyclopedia} of $I=\frac{1}{2}nc\varepsilon_{0}|E_{0}|^{2}<1.79074\times10^{17}$ W/m$^2$, where $n$ is the refractive index, $c$ is the speed of light in vacuum, and 
$\varepsilon_{0}$ is the vacuum permittivity. A refractive index $n\approx1.37$~\cite{wang2019broadband,yue2017photo,krishnamoorthy2020infrared} was observed in the Bi$_2$Se$_3$ (Bi$_2$Te$_3$) thin film under deep-ultraviolet frequency. Note that while a weaker laser intensity satisfies the high-frequency expansion more accurately, it also leads to a smaller Dirac gap, which may decrease experimental robustness.

\section{Time-reversal symmetry breaking with light}\label{Appendix_H}

Here, we prove that when $\varphi=0$, the Floquet Hamiltonian (\ref{eq:H_F}) in the main text still satisfies time-reversal symmetry. However, when $\varphi=\frac{\pi}{2}$, the time-reversal symmetry of the Hamiltonian (\ref{eq:H_F}) is broken.

The Floquet Hamiltonian~(\ref{eq:H_F}) under time-reversal transformation becomes
\begin{align}\label{eq:time_reversal}
&~~{\cal T}H^{(F)}({\bf k}){\cal T}^{-1} \nonumber\\
&\!=\!\sum_{j_z}C_{{\bf k},j_z}^{\dagger}\sigma_{0}\otimes\tau_{y}{\cal K}\left\{ \left[m_{0}-2t_{z}-4t_{||}+2{\cal J}_{0}(A(z)a)t_{||}\left[\cos(k_{x}a)+\cos(k_{y}a)\right]\right]\gamma_{0} \right.\nonumber\\
&\left.\!+\! {\cal J}_{0}(A_{0}a)\lambda_{||}\sin(k_{x}a)\gamma_{x}\!+\! {\cal J}_{0}(A(z)a)\lambda_{||}\sin(k_{y}a)\gamma_{y} \right\}{\cal K}^{-1}(\sigma_{0}\otimes\tau_{y})^{-1}C_{{\bf k},j_z} \nonumber\\
&\!+\!t_{z}\sum_{j_z}\!\left(\!C_{{\bf k},j_z}^{\dagger}\gamma_{0}C_{{\bf k},j_z+1} + C_{{\bf k},j_z+1}^{\dagger}\gamma_{0}C_{{\bf k},j_z} \!\right)\! 
\!-\! i\frac{\lambda_{z}}{2}\sum_{j_z}\!\left(\!C_{{\bf k},j_z}^{\dagger}\gamma_{z}C_{{\bf k},j_z+1}\!-\! C_{{\bf k},j_z+1}^{\dagger}\gamma_{z}C_{{\bf k},j_z}\!\right)\! \nonumber\\
&+\sum_{j_z}C_{{\bf k},j_z}^{\dagger}\sum_{n\in{\rm odd},n>0}\frac{2\lambda_{||}{\cal J}_{n}^{2}(A(z)a)}{n\hbar\omega}\sin(n\varphi)\sigma_{0}\otimes\tau_{y}{\cal K}\left\{ 2it_{||}\cos(k_{x}a)\sin(k_{y}a)[\gamma_{x}, \gamma_{0}] 
+2it_{||}\sin(k_{x}a)\cos(k_{y}a)[\gamma_{0}, \gamma_{y}] \right.\nonumber\\
&\left. - i\lambda_{||}\cos(k_{x}a)\cos(k_{y}a)[\gamma_{x}, \gamma_{y}] \right\}{\cal K}^{-1}(\sigma_{0}\otimes\tau_{y})^{-1}C_{{\bf k},j_z} \nonumber\\
&\!=\!\sum_{j_z}C_{{\bf k},j_z}^{\dagger}\left\{\left[m_{0}-2t_{z}-4t_{||}+2{\cal J}_{0}(A(z)a)t_{||}\left[\cos(k_{x}a)+\cos(k_{y}a)\right]\right]\gamma_{0} \right.\nonumber\\
&\left.\!-\! {\cal J}_{0}(A_{0}a)\lambda_{||}\sin(k_{x}a)\gamma_{x} \!-\! {\cal J}_{0}(A(z)a)\lambda_{||}\sin(k_{y}a)\gamma_{y} \right\}C_{{\bf k},j_z} \nonumber\\
&\!+\!t_{z}\sum_{j_z}\!\left(\!C_{{\bf k},j_z}^{\dagger}\gamma_{0}C_{{\bf k},j_z+1} + C_{{\bf k},j_z+1}^{\dagger}\gamma_{0}C_{{\bf k},j_z} \!\right)\! 
\!-\! i\frac{\lambda_{z}}{2}\sum_{j_z}\!\left(\!C_{{\bf k},j_z}^{\dagger}\gamma_{z}C_{{\bf k},j_z+1}\!-\! C_{{\bf k},j_z+1}^{\dagger}\gamma_{z}C_{{\bf k},j_z}\!\right)\! \nonumber\\
&\!+\!\sum_{j_z}C_{{\bf k},j_z}^{\dagger}\sum_{n\in{\rm odd},n>0}\frac{2i\lambda_{||}{\cal J}_{n}^{2}(A(z)a)}{n\hbar\omega}\sin(n\varphi)
\left\{ 2t_{||}\cos(k_{x}a)\sin(k_{y}a)[\gamma_{x}, \gamma_{0}] 
+2t_{||}\sin(k_{x}a)\cos(k_{y}a)[\gamma_{0}, \gamma_{y}] \right.\nonumber\\
&\left. + \lambda_{||}\cos(k_{x}a)\cos(k_{y}a)[\gamma_{x}, \gamma_{y}] \right\}C_{{\bf k},j_z} \nonumber\\
&\!=\! H^{(F)}(-{\bf k}) + \sum_{j_z}C_{{\bf k},j_z}^{\dagger}\sum_{n\in{\rm odd},n>0}\frac{4i\lambda_{||}{\cal J}_{n}^{2}(A(z)a)}{n\hbar\omega}\sin(n\varphi)
\left\{ 2t_{||}\cos(k_{x}a)\sin(k_{y}a)[\gamma_{x}, \gamma_{0}] 
+2t_{||}\sin(k_{x}a)\cos(k_{y}a)[\gamma_{0}, \gamma_{y}] \right.\nonumber\\
&\left. + \lambda_{||}\cos(k_{x}a)\cos(k_{y}a)[\gamma_{x}, \gamma_{y}] \right\}C_{{\bf k},j_z},
\end{align}
where ${\cal T}=\sigma_{y}\otimes\tau_{0}{\cal K}$~\cite{schindler2018higher} is the time-reversal operator with the complex conjugate operator ${\cal K}$, we have ${\cal K}H^{(F)}({\bf k}){\cal K}^{-1}=H^{(F)*}({\bf k})$, and we use
\begin{align}
H^{(F)}(-{\bf k}) 
&\!=\!\sum_{j_z}\!\left\{\!m_{0}-2t_{z}-4t_{||}+2{\cal J}_{0}(A(z)a)t_{||}\left[\cos(k_{x}a)+\cos(k_{y}a)\right]\!\right\}\!C_{{\bf k},j_z}^{\dagger}\gamma_{0}C_{{\bf k},j_z} \nonumber\\
&\!+\!t_{z}\sum_{j_z}\!\left(\!C_{{\bf k},j_z}^{\dagger}\gamma_{0}C_{{\bf k},j_z+1} + C_{{\bf k},j_z+1}^{\dagger}\gamma_{0}C_{{\bf k},j_z} \!\right)\! 
\!-\! i\frac{\lambda_{z}}{2}\sum_{j_z}\!\left(\!C_{{\bf k},j_z}^{\dagger}\gamma_{z}C_{{\bf k},j_z+1}\!-\! C_{{\bf k},j_z+1}^{\dagger}\gamma_{z}C_{{\bf k},j_z}\!\right)\! \nonumber\\
& \!-\! {\cal J}_{0}(A(z)a)\lambda_{||}\sin(k_{x}a)\sum_{j_z}C_{{\bf k},j_z}^{\dagger}\gamma_{x}C_{{\bf k},j_z} \!-\! {\cal J}_{0}(A(z)a)\lambda_{||}\sin(k_{y}a)\sum_{j_z}C_{{\bf k},j_z}^{\dagger}\gamma_{y}C_{{\bf k},j_z} \nonumber\\
&\!+\!\sum_{j_z}C_{{\bf k},j_z}^{\dagger}\sum_{n\in{\rm odd},n>0}\frac{2i\lambda_{||}{\cal J}_{n}^{2}(A(z)a)}{n\hbar\omega}\sin\left(\frac{n\pi}{2}\right)
\left\{ -2t_{||}\cos(k_{x}a)\sin(k_{y}a)[\gamma_{x}, \gamma_{0}] 
-2t_{||}\sin(k_{x}a)\cos(k_{y}a)[\gamma_{0}, \gamma_{y}] \right.\nonumber\\
&\left. - \lambda_{||}\cos(k_{x}a)\cos(k_{y}a)[\gamma_{x}, \gamma_{y}] \right\}C_{{\bf k},j_z}.
\end{align}
\begin{align}
&(\sigma_{y}\otimes\tau_{0})(\gamma_{0})(\sigma_{y}\otimes\tau_{0})^{-1}=\gamma_{0},\\
&(\sigma_{y}\otimes\tau_{0})(\gamma_{x})(\sigma_{y}\otimes\tau_{0})^{-1}=-\gamma_{x},\\
&(\sigma_{y}\otimes\tau_{0})({\cal K}\gamma_{y}{\cal K}^{-1})(\sigma_{y}\otimes\tau_{0})^{-1}=-\gamma_{y},\\
&(\sigma_{y}\otimes\tau_{0})(\gamma_{z})(\sigma_{y}\otimes\tau_{0})^{-1}=-\gamma_{z},\\
&(\sigma_{y}\otimes\tau_{0})[{\cal K}(i\gamma_{x}\gamma_{0}){\cal K}^{-1}](\sigma_{y}\otimes\tau_{0})^{-1}=i\gamma_{x}\gamma_{0},\\
&(\sigma_{y}\otimes\tau_{0})[{\cal K}(i\gamma_{0}\gamma_{x}){\cal K}^{-1}](\sigma_{y}\otimes\tau_{0})^{-1}=i\gamma_{0}\gamma_{x},\\
&(\sigma_{y}\otimes\tau_{0})[{\cal K}(i\gamma_{y}\gamma_{0}){\cal K}^{-1}](\sigma_{y}\otimes\tau_{0})^{-1}=i\gamma_{y}\gamma_{0},\\
&(\sigma_{y}\otimes\tau_{0})[{\cal K}(i\gamma_{0}\gamma_{y}){\cal K}^{-1}](\sigma_{y}\otimes\tau_{0})^{-1}=i\gamma_{0}\gamma_{y},\\
&(\sigma_{y}\otimes\tau_{0})[{\cal K}(i\gamma_{x}\gamma_{y}){\cal K}^{-1}](\sigma_{y}\otimes\tau_{0})^{-1}=-i\gamma_{x}\gamma_{y},\\
&(\sigma_{y}\otimes\tau_{0})[{\cal K}(i\gamma_{y}\gamma_{x}){\cal K}^{-1}](\sigma_{y}\otimes\tau_{0})^{-1}=-i\gamma_{y}\gamma_{x},
\end{align}
As a result of Eq.~(\ref{eq:time_reversal}), if $\varphi=0$, we have ${\cal T}H^{(F)}({\bf k}){\cal T}^{-1}=H^{(F)}(-{\bf k})$ which shows a time-reversal symmetry.
However, for $\varphi=\frac{\pi}{2}$, Eq.~(\ref{eq:time_reversal}) shows that the time-reversal symmetry is broken.

\section{Chern number}\label{Appendix_I}

Here, we use the definition of the Chern number~\cite{sun2020analytical,otrokov2019unique,fukui2005chern} on our specific model, and present some intermediate analytic simplifications that are useful for numerical implementation.
\begin{align}
C=\frac{1}{2\pi}\sum_{j}\int f(E_{j}-E_{F})\Omega_{j,z}(k_x,k_y)dk_{x}dk_{y},
\end{align} where $f(E_{j}-E_{F})$ is the Fermi distribution function, $f(E_{j}-E_{F})=\Theta(E_{F}-E_{j})$ and $\Theta(x)$ is the Heaviside step function (in the zero temperature limit)~\cite{qin2015three,qin2018high}, $E_{F}$ is the Fermi energy, and $\Omega_{j,z}(k_x,k_y)=\epsilon_{xyz}\Omega_{j,xy}(k_x,k_y)$ is the Berry curvature for the energy band $j$ with the three-component antisymmetric Levi-Civita tensor $\epsilon_{xyz}$, which reads~\cite{thouless1982quantized}
\begin{align}
\Omega_{j,xy}(k_x,k_y)
&=i\sum_{i\neq j}\frac{[\langle j|(\partial{\cal H}/\partial k_x)|i\rangle\langle i|(\partial{\cal H}/\partial k_y)|j\rangle - \langle j|(\partial{\cal H}/\partial k_y)|i\rangle\langle i|(\partial{\cal H}/\partial k_x)|j\rangle]}{(E_{j} - E_{i})^{2}} \nonumber\\
&=-{\rm Im}\sum_{i\neq j}\frac{[\langle j|(\partial{\cal H}/\partial k_x)|i\rangle\langle i|(\partial{\cal H}/\partial k_y)|j\rangle - \langle j|(\partial{\cal H}/\partial k_y)|i\rangle\langle i|(\partial{\cal H}/\partial k_x)|j\rangle]}{(E_{j} - E_{i})^{2}} \nonumber\\
&=-2{\rm Im}\sum_{i\neq j}\frac{\langle j|(\partial{\cal H}/\partial k_x)|i\rangle\langle i|(\partial{\cal H}/\partial k_y)|j\rangle}{(E_{j} - E_{i})^{2}}.
\end{align}
Here, the velocity operator $v_x$ along the $x$ direction is defined as $v_{x}=\partial{\cal H}/(\hbar\partial k_x)$ and the velocity operator $v_y$ along the $y$ direction is defined as $v_{y}=\partial{\cal H}/(\hbar\partial k_y)$.

Therefore, we have  
\begin{align}
\frac{\partial{\cal H}_{\rm tb}^{(0)}}{\partial k_x}=\begin{pmatrix}
\frac{\partial h}{\partial k_x} & 0 & 0 & \cdots & 0 \\
0 & \frac{\partial h}{\partial k_x} & 0 & \cdots & 0 \\
0 & 0 & \frac{\partial h}{\partial k_x} & \ddots & \vdots \\
\vdots& \ddots & \ddots & \ddots & 0 \\
0 & \cdots & 0 & 0 & \frac{\partial h}{\partial k_x}
\end{pmatrix}_{4N_z\times 4N_z},~~
\frac{\partial{\cal H}_{\rm tb}^{(0)}}{\partial k_y}=\begin{pmatrix}
\frac{\partial h}{\partial k_y} & 0 & 0 & \cdots & 0 \\
0 & \frac{\partial h}{\partial k_y} & 0 & \cdots & 0 \\
0 & 0 & \frac{\partial h}{\partial k_y} & \ddots & \vdots \\
\vdots& \ddots & \ddots & \ddots & 0 \\
0 & \cdots & 0 & 0 & \frac{\partial h}{\partial k_y}
\end{pmatrix}_{4N_z\times 4N_z},
\end{align} where
\begin{align}
\frac{\partial h}{\partial k_x}&\!=\!-2t_{||}a\sin(k_{x}a)\sigma_{0}\otimes\tau_{z} \!+\! \lambda_{||}a\cos(k_{x}a)\sigma_{x}\otimes\tau_{x}, \\
&\!=\!it_{||}a(e^{ik_{x}a}-e^{-ik_{x}a})\sigma_{0}\otimes\tau_{z} \!+\! \frac{\lambda_{||}a}{2}(e^{ik_{x}a}+e^{-ik_{x}a})\sigma_{x}\otimes\tau_{x}, \\
&\!=\!ia(T_{x}e^{ik_{x}a}-T_{x}^{\dagger}e^{-ik_{x}a}), \\
\frac{\partial h}{\partial k_y}&\!=\!-2t_{||}a\sin(k_{y}a)\sigma_{0}\otimes\tau_{z} \!+\! \lambda_{||}a\cos(k_{y}a)\sigma_{y}\otimes\tau_{x},\\
&\!=\!it_{||}a(e^{ik_{y}a}-e^{-ik_{y}a})\sigma_{0}\otimes\tau_{z} \!+\! \frac{\lambda_{||}a}{2}(e^{ik_{y}a}+e^{-ik_{y}a})\sigma_{y}\otimes\tau_{x}, \\
&\!=\!ia(T_{y}e^{ik_{y}a}-T_{y}^{\dagger}e^{-ik_{y}a}), \\
h&\!=\!\left[m_{0}-2t_{z}-4t_{||}\left(\sin^{2}\frac{k_{x}a}{2}+\sin^{2}\frac{k_{y}a}{2}\right) \right]\sigma_{0}\otimes\tau_{z} \!+\! \lambda_{||}\sin(k_{x}a)\sigma_{x}\otimes\tau_{x} \!+\! \lambda_{||}\sin(k_{y}a)\sigma_{y}\otimes\tau_{x} \nonumber\\
&\!=\!M_{0}+T_{x}e^{ik_{x}a}+T_{x}^{\dagger}e^{-ik_{x}a}+T_{y}e^{ik_{y}a}+T_{y}^{\dagger}e^{-ik_{y}a}.
\end{align} 
\begin{align}
M_{0}&\!=\!\left(m_{0}-2t_{z}-4t_{||}\right)\sigma_{0}\otimes\tau_{z},\\
T_{x}&\!=\!t_{||}\sigma_{0}\otimes\tau_{z}\!-\! i\frac{\lambda_{||}}{2}\sigma_{x}\otimes\tau_{x},~~
T_{x}^{\dagger}\!=\!t_{||}\sigma_{0}\otimes\tau_{z}\!+\! i\frac{\lambda_{||}}{2}\sigma_{x}\otimes\tau_{x},\\
T_{y}&\!=\!t_{||}\sigma_{0}\otimes\tau_{z}\!-\! i\frac{\lambda_{||}}{2}\sigma_{y}\otimes\tau_{x},~~
T_{y}^{\dagger}\!=\!t_{||}\sigma_{0}\otimes\tau_{z}\!+\! i\frac{\lambda_{||}}{2}\sigma_{y}\otimes\tau_{x},\\
T_{z}&\!=\!t_{z}\sigma_{0}\otimes\tau_{z}\!-\! i\frac{\lambda_{z}}{2}\sigma_{z}\otimes\tau_{x},~~
T_{z}^{\dagger}\!=\!t_{z}\sigma_{0}\otimes\tau_{z}\!+\! i\frac{\lambda_{z}}{2}\sigma_{z}\otimes\tau_{x}.
\end{align}

Furthermore, we have 
\begin{align}
\frac{\partial{\cal H}_{F}}{\partial k_x}=\begin{pmatrix}
\frac{\partial h_{F}}{\partial k_x} & 0 & 0 & \cdots & 0 \\
0 & \frac{\partial h_{F}}{\partial k_x} & 0 & \cdots & 0 \\
0 & 0 & \frac{\partial h_{F}}{\partial k_x} & \ddots & \vdots \\
\vdots& \ddots & \ddots & \ddots & 0 \\
0 & \cdots & 0 & 0 & \frac{\partial h_{F}}{\partial k_x}
\end{pmatrix}_{4N_z\times 4N_z},~~
\frac{\partial{\cal H}_{F}}{\partial k_y}=\begin{pmatrix}
\frac{\partial h_{F}}{\partial k_y} & 0 & 0 & \cdots & 0 \\
0 & \frac{\partial h_{F}}{\partial k_y} & 0 & \cdots & 0 \\
0 & 0 & \frac{\partial h_{F}}{\partial k_y} & \ddots & \vdots \\
\vdots& \ddots & \ddots & \ddots & 0 \\
0 & \cdots & 0 & 0 & \frac{\partial h_{F}}{\partial k_y}
\end{pmatrix}_{4N_z\times 4N_z},
\end{align} where
\begin{align}
\frac{\partial h_{F}}{\partial k_x}&\!=\!-2{\cal J}_{0}(A(z)a)t_{||}a\sin(k_{x}a)\gamma_{0} \!+\! {\cal J}_{0}(A(z)a)\lambda_{||}a\cos(k_{x}a)\gamma_{x} \nonumber\\
&\!+\!\sum_{n\in{\rm odd},n>0}\frac{2i\lambda_{||}{\cal J}_{n}^{2}(A(z)a)}{n\hbar\omega}\sin\left(n\varphi\right)
\left\{ -2t_{||}a\sin(k_{x}a)\sin(k_{y}a)[\gamma_{x}, \gamma_{0}] 
+2t_{||}a\cos(k_{x}a)\cos(k_{y}a)[\gamma_{0}, \gamma_{y}] \right.\nonumber\\
&\left. + \lambda_{||}a\sin(k_{x}a)\cos(k_{y}a)[\gamma_{x}, \gamma_{y}] \right\},\\
\frac{\partial h_{F}}{\partial k_y}&\!=\!-2{\cal J}_{0}(A(z)a)t_{||}a\sin(k_{y}a)\gamma_{0} \!+\! {\cal J}_{0}(A(z)a)\lambda_{||}a\cos(k_{y}a)\gamma_{y} \nonumber\\
&\!+\!\sum_{n\in{\rm odd},n>0}\frac{2i\lambda_{||}{\cal J}_{n}^{2}(A(z)a)}{n\hbar\omega}\sin\left(n\varphi\right)
\left\{ 2t_{||}a\cos(k_{x}a)\cos(k_{y}a)[\gamma_{x}, \gamma_{0}] 
-2t_{||}a\sin(k_{x}a)\sin(k_{y}a)[\gamma_{0}, \gamma_{y}] \right.\nonumber\\
&\left. + \lambda_{||}a\cos(k_{x}a)\sin(k_{y}a)[\gamma_{x}, \gamma_{y}] \right\},\\
h_{F}&\!=\! \left\{\!m_{0}-2t_{z}-4t_{||}+2{\cal J}_{0}(A(z)a)t_{||}\left[\cos(k_{x}a)+\cos(k_{y}a)\right]\!\right\}\gamma_{0} \!+\! {\cal J}_{0}(A(z)a)\lambda_{||}\sin(k_{x}a)\gamma_{x} \!+\! {\cal J}_{0}(A(z)a)\lambda_{||}\sin(k_{y}a)\gamma_{y} \nonumber\\
&\!+\!\sum_{n\in{\rm odd},n>0}\frac{2i\lambda_{||}{\cal J}_{n}^{2}(A(z)a)}{n\hbar\omega}\sin\left(n\varphi\right)
\left\{ 2t_{||}\cos(k_{x}a)\sin(k_{y}a)[\gamma_{x}, \gamma_{0}] 
+2t_{||}\sin(k_{x}a)\cos(k_{y}a)[\gamma_{0}, \gamma_{y}] \right.\nonumber\\
&\left. - \lambda_{||}\cos(k_{x}a)\cos(k_{y}a)[\gamma_{x}, \gamma_{y}] \right\}.
\end{align} 

\clearpage

\section{Localization of the bottom surface state}\label{Appendix_J}

The motivation of this Appendix \ref{Appendix_J} is to discuss the localization of the bottom surface state with only magnetic doping.

\begin{figure}
\centering
\includegraphics[width=0.8\textwidth]{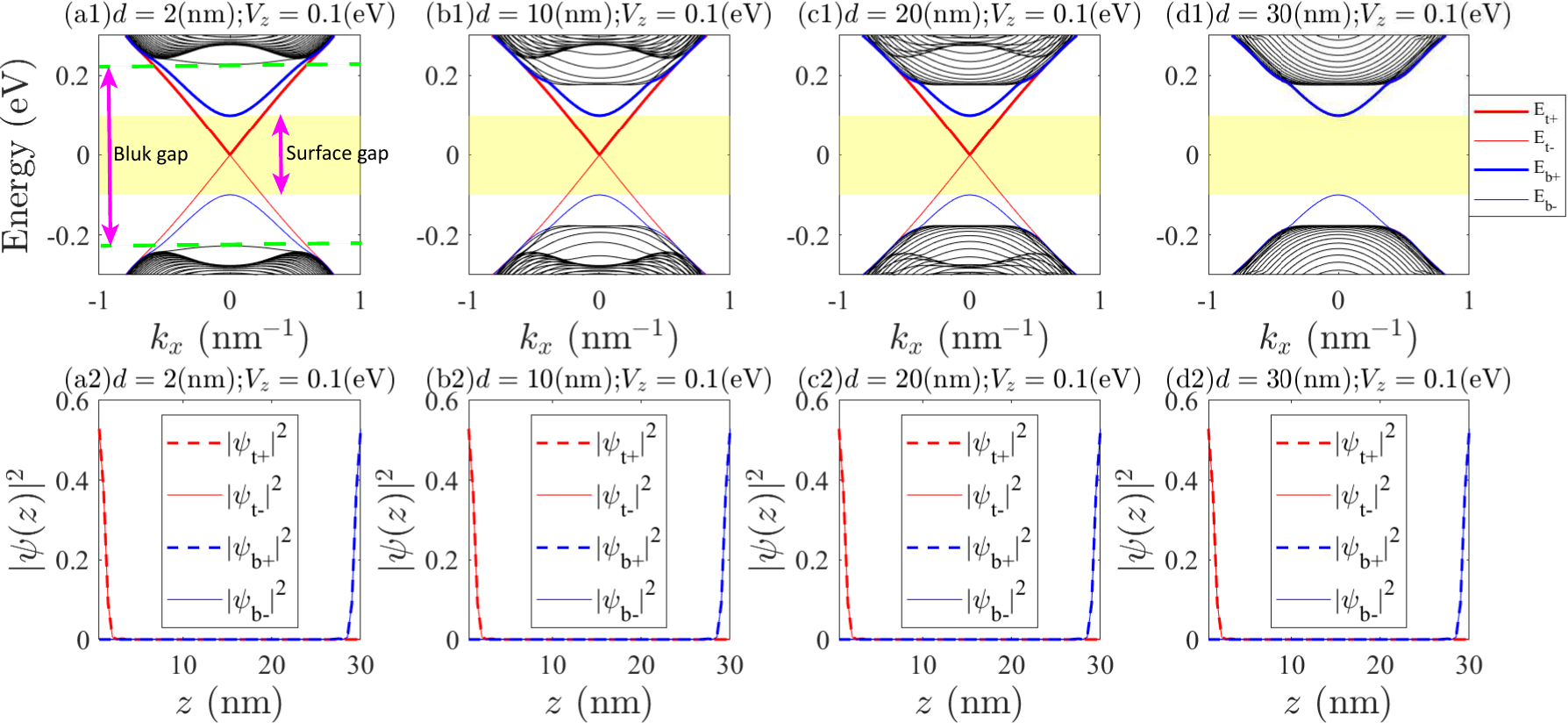}
\caption{(Color online)
[(a1)-(d1)] Band structures with only magnetic doping and $k_y=0$ for (a1) $d=2$ nm; (b1) $d=10$ nm; (c1) $d=20$ nm; (d1) $d=30$ nm. Here, ``$d$'' is the penetration depth of Cr doping into the bulk. 
$E_{t\pm}$ and $E_{b\pm}$ are the energies of the top and bottom surface bands; the subscripts ``$\pm$'' respectively denote the lowest conduction band or highest valence band. 
The yellow shaded intervals indicate the band widths of the bottom surface. 
Indeed, the bulk gap is always much larger than the bottom surface gap.
The total length along $z$ direction is $L_z=30$ nm. Therefore, when $d=L_z=30$ nm, both the top and bottom surfaces are gapped, as shown in (d1).
[(a2)-(d2)] Spatial state distribution $|\psi(z)|^{2}$  with (a2) $d=2$ nm, (b2) $d=10$ nm, (c2) $d=20$ nm, (d2) $d=30$ nm, along the vertical direction $z$ for the lowest conduction (subscript ``+\rq\rq) and highest valence (subscript ``-\rq\rq) bands of the top surface state ($|\psi_{t\pm}|^{2}$ -- red lines) and bottom surface state ($|\psi_{b\pm}|^{2}$ -- blue lines) at $k_{x}=k_{y}=0$. Indeed, the supposed top and bottom surface states are concentrated near the top ($z=0.5$ nm) and bottom $z=30$ nm boundaries. 
The other parameters are $A_{0}=0$, $V_{z}=0.1$ eV, $L_z=30$ nm, $a_{z}=0.5$ nm, $a=1$ nm, $t_{z}=0.40$ eV, $t_{||}=0.566$ eV, $\lambda_{z}=0.44$ eV, $\lambda_{||}=0.41$ eV, and $m_{0}=0.28$ eV.}
\label{Fig:E_P_Lz30_reply1_together}
\end{figure}

A topological insulator has a topologically protected bulk gap but gapless Dirac cones localized on the top and
bottom surfaces. When the local-time reversal symmetry is broken by the magnetic
doping, the gapless surface Dirac cone will open an energy gap, accompanied by a
half-quantized surface chiral Hall current. This surface energy gap is determined by
the magnitude of the magnetic doping. Therefore, if the magnitude of the magnetic
doping is not strong enough to flip the bulk energy bands, the localization of the
bottom surface state will not be significantly affected by the extent of magnetic
doping penetration into the bulk material. In this case, even if magnetization enters
the bulk, it will only induce a modification of the bulk energy gap, and the bulk of the
system will remain insulating, having little impact on the surface states. Furthermore, we will confirm this conclusion through numerical calculations.

We set $d$ as the penetration depth of Cr doping into the bulk. Without light, if $V_{z}(j)$ penetrates into the bulk, for example, $d=2$ nm, $d=10$ nm, $d=20$ nm, $d=30$ nm (compared to $d=1$ nm in the main text), the wavefunction will still localize on the bottom surface as shown in Fig.~\ref{Fig:E_P_Lz30_reply1_together}(a2)--\ref{Fig:E_P_Lz30_reply1_together}(d2).

As shown in upper line (band structures) of Fig.~\ref{Fig:E_P_Lz30_reply1_together}(a1)--\ref{Fig:E_P_Lz30_reply1_together}(d1), the chosen magnitude of the magnetic impurity is 0.1 eV, i.e., the bottom surface gap (which is represented by the yellow shaded intervals) is about 0.2 eV, which is always smaller than the bulk gap [which is delineated between two dashed green lines as shown in Fig.~\ref{Fig:E_P_Lz30_reply1_together}(a1) and its counterparts] with different penetration depth of Cr doping. Therefore, the wavefunction of the bottom surface state always localizes on the bottom surface as shown in the lower line (spatial state distribution) of Fig.~\ref{Fig:E_P_Lz30_reply1_together}(a2)--\ref{Fig:E_P_Lz30_reply1_together}(d2).

Particularly when $d=L_z=30$ nm, both the top and bottom surfaces are gapped, as shown in Fig.~\ref{Fig:E_P_Lz30_reply1_together}(d1). Here, the total length along the $z$ direction is $L_z=30$ nm.

\section{Semi-Floquet phase to Floquet Chern insulator phase}\label{Appendix_K}

The motivation of this Appendix \ref{Appendix_K} is to discuss the crossover from the semi-Floquet phase to the Floquet Chern insulator phase with only light pumping.

\begin{figure}
\centering
\includegraphics[width=0.85\textwidth]{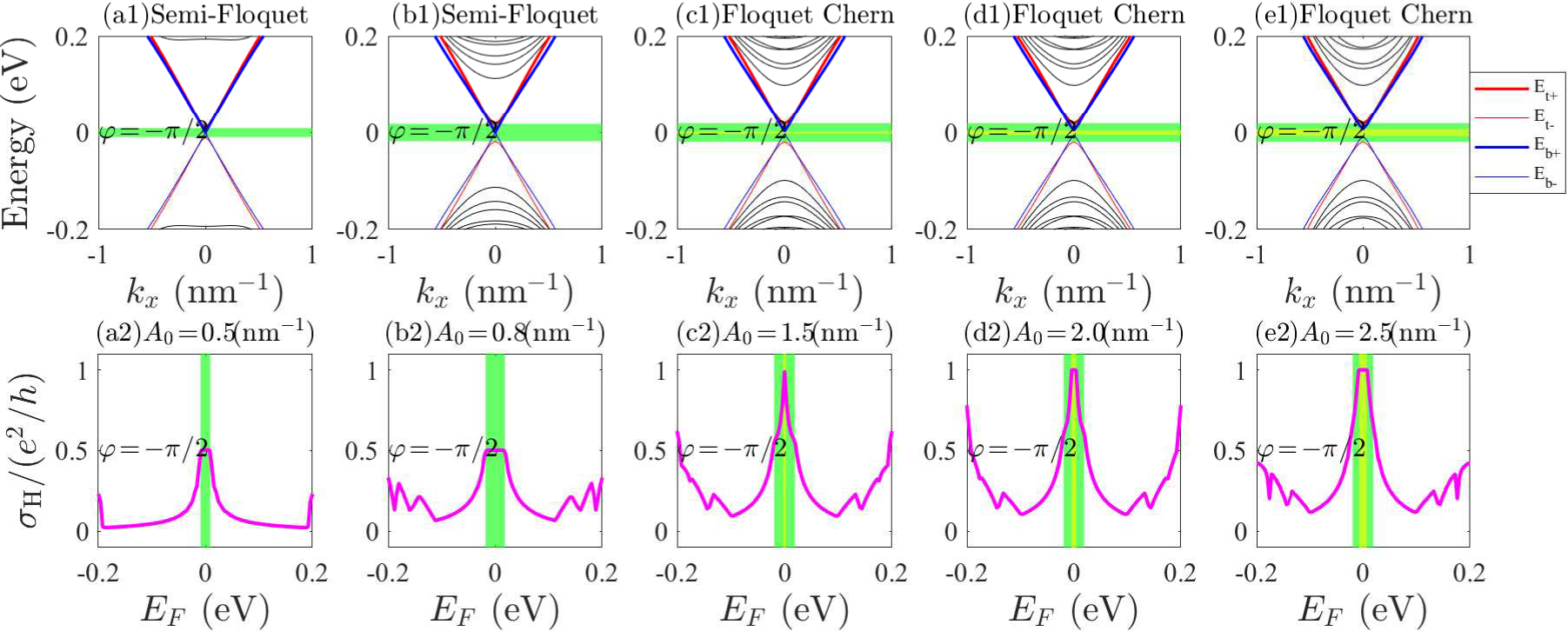}
\caption{(Color online)
[(a1)-(e1)] Band structures with only light pumping and $k_y=0$ for (a1) $A_{0}=0.5$ nm$^{-1}$; (b1) $A_{0}=0.8$ nm$^{-1}$; (c1) $A_{0}=1.5$ nm$^{-1}$; (d1) $A_{0}=2.0$ nm$^{-1}$; (e1) $A_{0}=2.5$ nm$^{-1}$.
Here, $E_{t\pm}$ and $E_{b\pm}$ are the energies of the top and bottom surface bands; the subscripts ``$\pm$'' respectively denote the lowest conduction band or highest valence band. 
The green and yellow shaded intervals indicate the band widths of the top and bottom surfaces, respectively. 
[(a2)-(e2)] Hall conductivities with (a2) $A_{0}=0.5$ nm$^{-1}$; (b2) $A_{0}=0.8$ nm$^{-1}$; (c2) $A_{0}=1.5$ nm$^{-1}$; (d2) $A_{0}=2.0$ nm$^{-1}$; (e2) $A_{0}=2.5$ nm$^{-1}$. 
The other parameters are $V_{z}=0$, $L_z=30$ nm, $a_{z}=0.5$ nm, $a=1$ nm, $t_{z}=0.40$ eV, $t_{||}=0.566$ eV, $\lambda_{z}=0.44$ eV, $\lambda_{||}=0.41$ eV, $m_{0}=0.28$ eV, $\delta=16.3$ nm, $\hbar\omega=3.82$ eV, and $\varphi=-\pi/2$.}
\label{Fig:E_C_Lz30_delta163_inverse_reply1_together}
\end{figure}

With continuously tuned light intensity, we plot the crossover from the semi-Floquet phase to the Floquet Chern insulator phase as shown in Fig.~\ref{Fig:E_C_Lz30_delta163_inverse_reply1_together}.

As shown in Figs.~\ref{Fig:E_C_Lz30_delta163_inverse_reply1_together}(a1) and \ref{Fig:E_C_Lz30_delta163_inverse_reply1_together}(b1) for the semi-Floquet case, the light is coming from the top surface with a weak light intensity, such that only the top surface (red curves) opens up a gap and the bottom surface (blue curves) is gapless. Only the gapped top surface contributes a half-quantized Hall conductance within its gap (green shaded interval), as shown in Figs.~\ref{Fig:E_C_Lz30_delta163_inverse_reply1_together}(a2) and \ref{Fig:E_C_Lz30_delta163_inverse_reply1_together}(b2). In Figs.~\ref{Fig:E_C_Lz30_delta163_inverse_reply1_together}(c1), \ref{Fig:E_C_Lz30_delta163_inverse_reply1_together}(d1), and \ref{Fig:E_C_Lz30_delta163_inverse_reply1_together}(e1) for the Floquet Chern case, the light is coming from the top surface with a strong light intensity, such that it penetrates both the top and bottom surfaces and gaps out their Dirac cones, which together contribute a quantized Hall conductivity, as shown in the gapped region of both surfaces (yellow shaded interval) in Figs.~\ref{Fig:E_C_Lz30_delta163_inverse_reply1_together}(c2), \ref{Fig:E_C_Lz30_delta163_inverse_reply1_together}(d2), and \ref{Fig:E_C_Lz30_delta163_inverse_reply1_together}(e2).

\section{Quench dynamics from axion insulator phase to Chern insulator phase}\label{Appendix_L}

The motivation of this Appendix \ref{Appendix_L} is to discuss the crossover from the axion insulator phase to the Chern insulator phase with a continuously tuned time duration parameter $T_1$.

\begin{figure}
\centering
\includegraphics[width=0.9\textwidth]{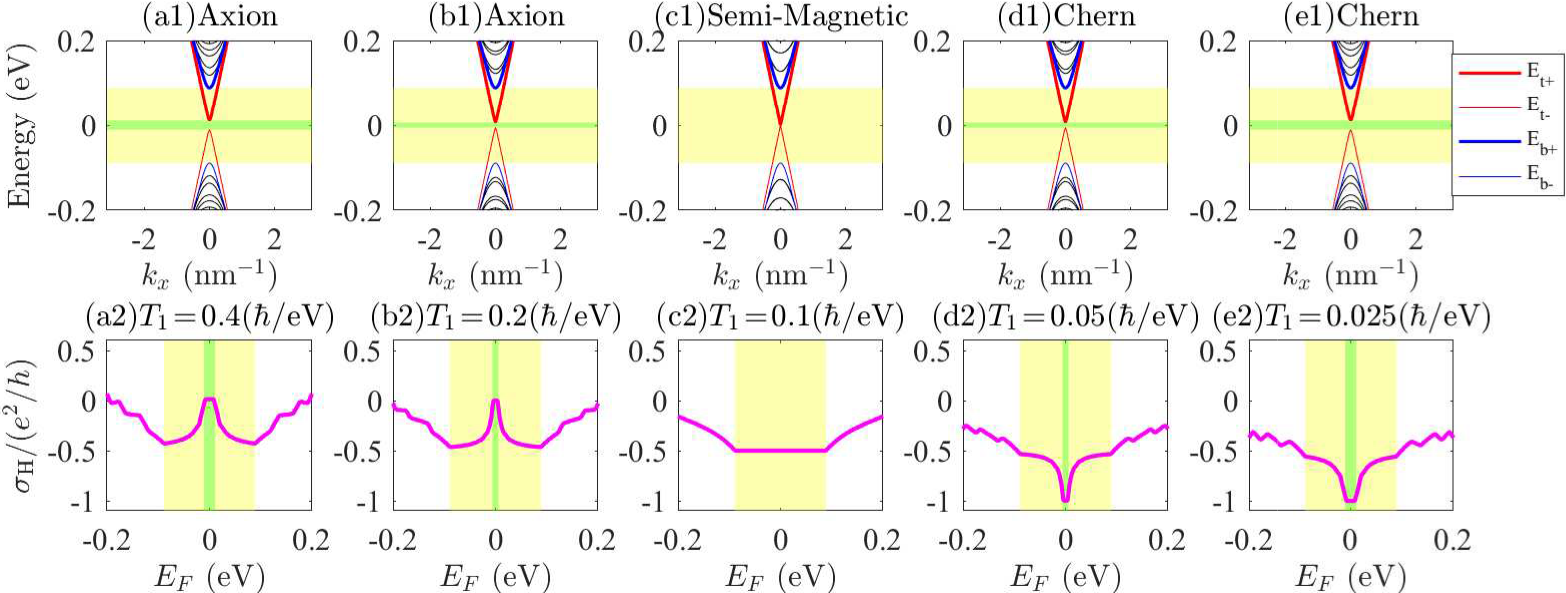}
\caption{(Color online)
[(a1)-(e1)] Band structures with magnetic doping ($V_z=0.1$ eV), weak intensity of light ($A_{0}=0.8$ nm$^{-1}$), and fixed $T_{2}=0.1$ $\hbar/$eV for (a1) $T_{1}=0.4$ $\hbar/$eV; (b1) $T_{1}=0.2$ $\hbar/$eV; (c1) $T_{1}=0.1$ $\hbar/$eV; (d1) $T_{1}=0.05$ $\hbar/$eV; (e1) $T_{1}=0.025$ $\hbar/$eV.
Here, $E_{t\pm}$ and $E_{b\pm}$ are the energies of the top and bottom surface bands; the subscripts ``$\pm$'' respectively denote the lowest conduction band or highest valence band. 
The green and yellow shaded intervals indicate the band widths of the top and bottom surfaces, respectively. 
[(a2)-(e2)] Hall conductivities with (a2) $T_{1}=0.4$ $\hbar/$eV; (b2) $T_{1}=0.2$ $\hbar/$eV; (c2) $T_{1}=0.1$ $\hbar/$eV; (d2) $T_{1}=0.05$ $\hbar/$eV; (e2) $T_{1}=0.025$ $\hbar/$eV. 
The other parameters are $d=1$ nm, $L_z=30$ nm, $a_{z}=0.5$ nm, $a=1$ nm, $t_{z}=0.40$ eV, $t_{||}=0.566$ eV, $\lambda_{z}=0.44$ eV, $\lambda_{||}=0.41$ eV, $m_{0}=0.28$ eV, $\delta=16.3$ nm, $\hbar\omega=3.82$ eV, and $\varphi=\pi/2$.}
\label{Fig:Quench_E_C_Lz30_reply1_together}
\end{figure}

With fixed $T_2$ and continuously tuned $T_1$, we plot the crossover from the axion insulator phase to the Chern insulator phase as shown in Fig.~\ref{Fig:Quench_E_C_Lz30_reply1_together}.

As shown in Figs.~\ref{Fig:Quench_E_C_Lz30_reply1_together}(a1)-\ref{Fig:Quench_E_C_Lz30_reply1_together}(e1), one can find that the value of the gap of the top surface (red curve) can be tuned by the time duration parameter $T_1$. When $T_1>T_2$, the system is in the axion insulator phase, as shown in Figs.~\ref{Fig:Quench_E_C_Lz30_reply1_together}(a2) and \ref{Fig:Quench_E_C_Lz30_reply1_together}(b2). When $T_1<T_2$, the system is in the Chern insulator phase, as shown in Figs.~\ref{Fig:Quench_E_C_Lz30_reply1_together}(d2) and \ref{Fig:Quench_E_C_Lz30_reply1_together}(e2). Particularly, when $T_1=T_2$, as shown in Figs.~\ref{Fig:Quench_E_C_Lz30_reply1_together}(c1) and \ref{Fig:Quench_E_C_Lz30_reply1_together}(c2), the system becomes a semi-magnetic topological insulator phase, which is different from the original axion insulator and Chern insulator phases.

\section{Phases under a strong intensity of light}\label{Appendix_M}

The motivation of this Appendix \ref{Appendix_M} is to investigate the phases under a strong intensity of light.

\begin{figure}
\centering
\includegraphics[width=0.4\textwidth]{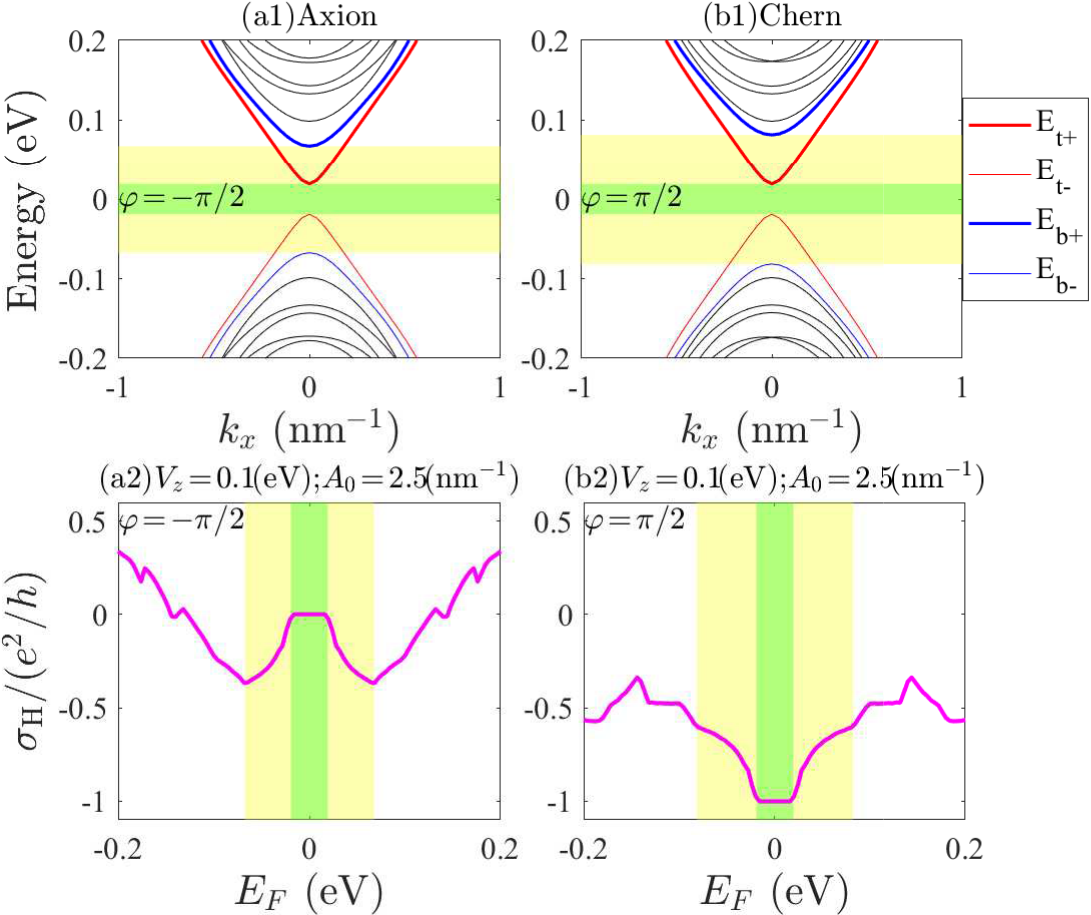}
\caption{(Color online)
[(a1)-(b1)] Band structures with magnetic doping ($V_z=0.1$ eV) and strong intensity of light ($A_{0}=2.5$ nm$^{-1}$) for (a1) $\varphi=-\pi/2$ (left-handed polarization) and (b1) $\varphi=\pi/2$ (right-handed polarization).
Here, $E_{t\pm}$ and $E_{b\pm}$ are the energies of the top and bottom surface bands; the subscripts ``$\pm$'' respectively denote the lowest conduction band or highest valence band. 
The green and yellow shaded intervals indicate the band widths of the top and bottom surfaces, respectively. 
[(a2)-(b2)] Hall conductivities with (a2) $\varphi=-\pi/2$ (left-handed polarization) and (b2) $\varphi=\pi/2$ (right-handed polarization). 
The other parameters are $L_z=30$ nm, $a_{z}=0.5$ nm, $a=1$ nm, $t_{z}=0.40$ eV, $t_{||}=0.566$ eV, $\lambda_{z}=0.44$ eV, $\lambda_{||}=0.41$ eV, $m_{0}=0.28$ eV, $d=1$ nm, $\delta=16.3$ nm, and $\hbar\omega=3.82$ eV.}
\label{Fig:E_C_Vz01_A25_Lz30_delta163_together}
\end{figure}

We plot the energy spectrum and the Hall conductivities with opposite polarized optical chirality under a strong intensity of light ($A_0=2.5$ nm$^{-1}$) as shown in Fig.~\ref{Fig:E_C_Vz01_A25_Lz30_delta163_together}.

As shown in Figs.~\ref{Fig:E_C_Vz01_A25_Lz30_delta163_together}(a1) and \ref{Fig:E_C_Vz01_A25_Lz30_delta163_together}(b1), both the top (red) and bottom (blue) surface Dirac cones are gapped. With left-handed polarization, the Dirac cones are of opposite chirality, resulting in opposite Hall conductivities that cancel within the gap (green), as shown in Fig.~\ref{Fig:E_C_Vz01_A25_Lz30_delta163_together}(a2). With right-handed polarization, the top surface Dirac cone's chirality is flipped, giving rise to an integer quantized Hall conductivity within the gap (green), as shown in Fig.~\ref{Fig:E_C_Vz01_A25_Lz30_delta163_together}(b2). These are, respectively, the axion and Chern insulators, which are very similar to those in the cases that are under a weak intensity of light ($A_0=0.8$ nm$^{-1}$) as shown in Figs. \ref{Fig:E_C_Lz30_delta163_inverse_together}(b1, c1) and \ref{Fig:E_C_Lz30_delta163_inverse_together}(b2, c2) in the main text.

Different from the case of weak light intensity, the band width of the bottom surface (indicated by the yellow shaded interval) undergoes a slight modification under the influence of strong light intensity.

\bibliography{references_Floquet_Axion}

\end{document}